\pdfoutput=1
\documentclass[man, 12pt, floatsintext]{apa7}
\usepackage[american]{babel}
\usepackage{graphicx}
\usepackage[natbibapa]{apacite}
\makeatletter\AtBeginDocument{\NAT@longnamesfalse}\makeatother
\usepackage{pdfpages}
\usepackage{makecell}
\usepackage{amsthm}
\usepackage{arydshln}
\usepackage{array}
\usepackage{algorithm}
\usepackage{threeparttable}
\usepackage{booktabs}
\usepackage[noEnd]{algpseudocodex}
\usepackage{multirow}
\usepackage{subfig}
\usepackage{bm}
\usepackage{siunitx}
\usepackage{tabularray}
\usepackage{lscape}
\usepackage{pdflscape}
\usepackage{rotating}
\usepackage{setspace}
\usepackage{caption}
\usepackage{mathtools}
\usepackage{subfig}
\usepackage{amsfonts}
\usepackage{amsmath}
\usepackage[skip=2pt]{caption}
\usepackage{enumerate}
\usepackage{enumitem}
\usepackage{float}
\restylefloat{table}
\usepackage{indentfirst}
\usepackage{parskip}
\usepackage{amssymb}
\usepackage{fancyhdr}
\usepackage{multirow}
\usepackage{longtable}
\usepackage{verbatim}
\usepackage{fixdif}
\usepackage[most]{tcolorbox}
\usepackage{tabularx}
\usepackage{etoolbox}
\AtBeginDocument{%
  \appto\normalsize{%
    \abovedisplayskip=4pt plus 2pt minus 2pt
    \belowdisplayskip=4pt plus 2pt minus 2pt
    \abovedisplayshortskip=0pt plus 2pt
    \belowdisplayshortskip=4pt plus 2pt minus 2pt}%
  \normalsize}
\usepackage{listings}
\lstdefinestyle{rcode}
{language=R,basicstyle=\ttfamily\small,keywordstyle=\color{blue!70!black},commentstyle=\color{gray},stringstyle=\color{teal!70!black},frame=single,rulecolor=\color{black!30},backgroundcolor=\color{gray!5},showstringspaces=false,breaklines=true,numbers=none,xleftmargin=0.5em,framexleftmargin=0.5em,framexrightmargin=0.5em,aboveskip=8pt,belowskip=8pt}
\makeatletter
\newcommand{\nextverbatimspread}[1]{%
  \def\verbatim@font{%
    \linespread{#1}\normalfont\ttfamily%
    \gdef\verbatim@font{\normalfont\ttfamily}}%
}

\fancypagestyle{titlepage}{\fancyhf{}}
\newcommand\independent{\protect\mathpalette{\protect\independenT}{\perp}}
\def\independenT#1#2{\mathrel{\rlap{$#1#2$}\mkern2mu{#1#2}}}

\newcommand\notsotiny{\@setfontsize\notsotiny\@vipt\@viipt}
\newcommand{\RNum}[1]{\uppercase\expandafter{\romannumeral #1\relax}}

\newtheorem{lem}{Lemma}
\newtheorem{proposition}{Proposition}
\newcommand{\Rmnum}[1]{\expandafter\@slowromancap\romannumeral #1@}
\makeatother

\bmdefine{\w}{w}
\bmdefine{\x}{x}
\bmdefine{\g}{g}

\def\bs{\boldsymbol}

\usepackage[dvipsnames]{xcolor}

\shorttitle{Bayesian Hierarchical Models for Organizational Effectiveness}
\title{Evaluating Organizational Effectiveness in Multi-Site Randomized Trials via Bayesian Hierarchical Modeling}
\authorsnames[1,1,2,3,3,1]{Mingya Huang, Xinran Li, Zhengyan Xu, Jonah Deutsch, Peter Kress, Guanglei Hong}
\authorsaffiliations{{University of Chicago}, {University of Pennsylvania}, {Mathematica}}
\leftheader{Huang, Li, Xu, Deutsch, Kress, \& Hong}
\abstract{Evaluating the relative effectiveness of organizations such as job training centers is essential for identifying exemplary practices and targeting capacity-building support. A multi-site randomized trial identifies the intention-to-treat effect of the program relative to the control condition at each site. However, comparing the ITT effects across sites ignores differences in local ecological conditions. We define local relative effectiveness as the program impact at a site solely attributable to malleable organizational practices, relative to peer organizations operating under similar local conditions. In multi-site randomized trials, we identify it as the part of a site’s ITT effect remaining after adjusting for baseline covariates and the site's control group outcome, a proxy for unmeasured pre-existing contextual conditions. This study addresses limitations of existing procedures in estimation and inference, which include finite sample bias when estimated site-level quantities are taken as known predictors and outcomes with correlated errors. We propose Bayesian hierarchical models that jointly estimate site-level parameters and their cross-site distribution, with estimation uncertainty propagating into the results. Simulations calibrated to the National Job Corps Study show reduced root mean squared error, especially with small within-site samples, and near-nominal credible interval coverage. A reanalysis of the application data estimates each Job Corps center’s relative effectiveness and posterior probabilities of lower, middle, or upper tier membership, informing contract renewals and oversight.}
\authornote{\addORCIDlink{Mingya Huang}{0000-0002-0647-7390}
Correspondence concerning this article should be addressed to Mingya Huang, Department of Comparative Human Development, University of Chicago, IL 60637. E-mail: mhuang233@uchicago.edu}
\begin{document}
\setcounter{page}{0}
\maketitle

\section{Introduction}

Intervention programs in each field are often delivered by multiple organizations, such as workforce training centers, public schools, or health care clinics. Organizational theory recognizes effectiveness as multidimensional, with its assessment depending on the criteria and reference points used. Understanding variation in effectiveness across organizations implementing the same program is an important subject of study \citep{cameron1983organizational}. Such comparisons inform decisions about contract renewal, provider rankings, and improvement efforts. Their validity depends on how organizational effectiveness is measured, a central concern in nonprofit and public management accountability and the study of exemplary practices \citep{lee2015framework}.

Multi-site randomized trials provide an opportunity to evaluate organizational effectiveness in program delivery. Within each site, eligible participants are randomized to treatment (program access) or control (no access). The site-specific intention-to-treat (ITT) effect is identified by the difference in mean outcomes between the two groups. However, differences in ITT effects across sites may not reflect differences in organizational effectiveness because sites operate under different local conditions. Participant composition, alternative programs available to control participants, and other contextual factors can influence ITT effects independently of organizational practice \citep{weiss2014conceptual}. For accountability and improvement, the goal is to distinguish the contribution of organizational practice from these local conditions.

Separating organizational effectiveness from local conditions is a longstanding challenge. A common approach adjusts for observed characteristics and attributes the remaining variation in outcomes to organizations. In education, value-added models regress student outcomes on baseline characteristics and interpret residual differences across schools or teachers as differences in effectiveness \citep{Rubin_2004, Raudenbush_2004, Chetty_2014}. Related work studies the reassignment of teachers using experimental data \citep{graham2023teacher}, compares organizations with the best performers in the sample \citep{horrace2023lasso}, and characterizes the distribution of unobserved organizational effects \citep{bonhomme2022posterior}. In multi-site trials, \citet{Lu_2023} use balancing weights to compare sites serving similar participants. These approaches adjust only for observed baseline conditions, so unmeasured differences between sites remain confounded with organizational effectiveness. Multi-site randomized trials help address this limitation. Because of within-site randomization, control-group outcomes reflect local conditions, including those not fully captured by observed covariates.

Building on prior work on organizational effectiveness in multi-site randomized trials \citep{Hong2026AJE}, we develop Bayesian hierarchical models to estimate each site's local relative effectiveness (LRE), addressing challenges in both estimation and inference. LRE is defined as the program impact at a site not explained by local conditions and thus attributed to organizational practice. Our models treat the site-level quantities as unknown parameters and estimate them jointly with their between-site distribution. This avoids the resulting bias in LRE estimates and carries the estimation error into the credible intervals and tier probabilities.

\subsection{Motivating Application: the National Job Corps Study (NJCS)}
\label{sec:motivation}

The NJCS illustrates the problem concretely. This federal evaluation randomized disadvantaged youth ages 16 to 24 to Job Corps enrollment or a control condition at about 100 centers across the United States \citep{Schochet_2008}. The centers operate under widely varying local conditions, such as labor markets, alternative training options, and applicant composition, some of which are hard to measure at baseline.

\begin{figure}[!htbp]
{\centering
\includegraphics[width=0.85\linewidth]{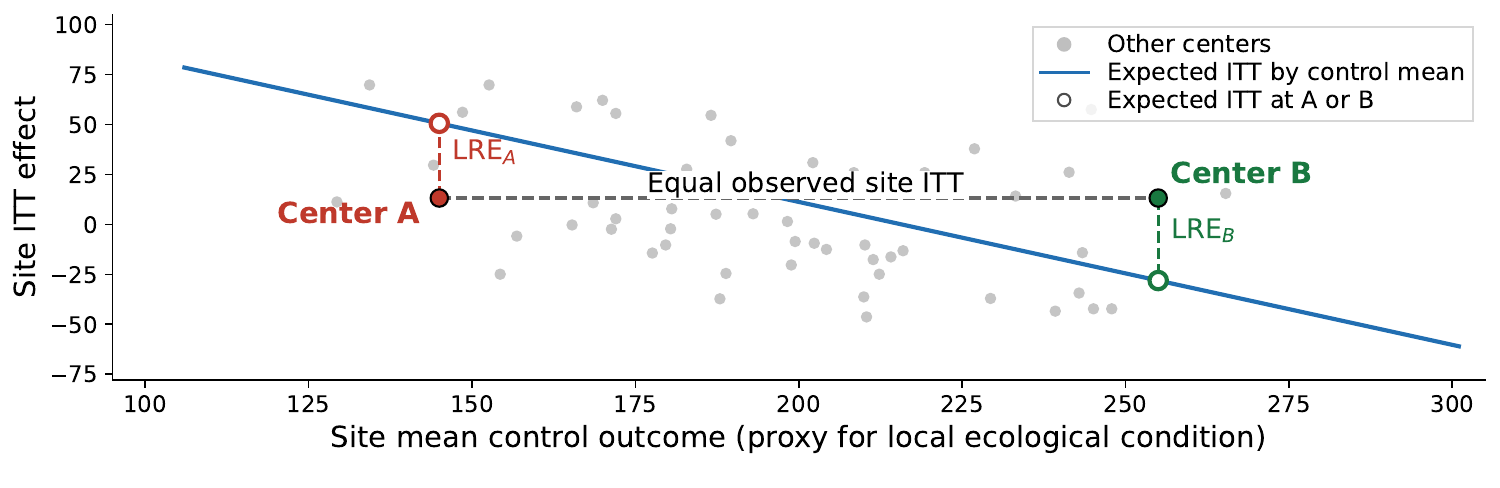}\par}
\caption{Local Relative Effectiveness in a Hypothetical Multi-site Trial. Each dot represents a center's ITT effect versus its mean control outcome. The solid line shows the conditional mean ITT effect. LRE is the signed vertical distance from this line, positive above and negative below. Centers A and B have equal ITT effects but opposite-signed LREs.}
\label{fig:motivation}
\end{figure}

\citet{Hong2026AJE} find that centers with higher average control-group outcomes tend to have smaller program impacts, even after adjusting for observed site-level characteristics. Figure~\ref{fig:motivation} illustrates this pattern using two hypothetical centers, A and B, with identical site-specific ITT effects but different local conditions proxyed by their control-group mean outcomes. To assess their relative effectiveness, we compare each center's ITT effect with the average ITT effect among centers with similar control mean outcomes. Center A falls below its benchmark, while Center B exceeds it. In the absence of unmeasured confounding, these deviations represent their LREs, reflecting organizational effectiveness relative to centers under similar local conditions \citep{Hong2026AJE}. This motivates the central question of how to estimate LRE accurately.

\subsection{Current Challenges}
\label{sec:challenges}

To account for differences in local conditions, \citet{Hong2026AJE} adjust for site-specific control-group mean outcomes in addition to observed baseline characteristics. Their two-step mixed-effects (2S-ME) procedure estimates in Step 1 each site's control-group mean outcome with empirical Bayes shrinkage \citep{Morris_1983}, which pulls noisy estimates toward the grand mean. In Step 2, it fits a mixed-effects model with site-specific random treatment effects, adjusting for the shrunken control means and observed site-level characteristics. Each site's LRE is the empirical Bayes estimate of its adjusted random treatment effect. In their simulations, this approach reduces bias in LRE estimates compared with methods that adjust only for observed baseline covariates.

However, two challenges remain. The first concerns measurement error. Each site's control-group mean outcome and covariate means are estimated from finite samples, but the Step-2 regression treats them as error-free, leading to errors-in-variables (EIV) bias, which we call plug-in bias in this study. Moreover, the control mean enters the Step-2 regression twice: as a predictor, and with a negative sign in the outcome, since the ITT estimate is the treated mean minus the control mean. Its sampling error is therefore negatively correlated across the outcome and the predictor, which adds a further bias to the slope beyond classical attenuation. Clinical trial meta-analysis has addressed this problem for the relationship between treatment effects and baseline risk, mainly with Bayesian models \citep{Thompson_1997}. Our target is different. The LRE is the residual from the Step-2 regression, so error in the site-level summaries both biases the fitted adjustment and carries into the residual.

The second challenge concerns statistical inference. Treating estimated site-level summaries as known can lead to poorly calibrated intervals for LREs. Empirical Bayes intervals may also under-cover when uncertainty in the estimated between-site distribution is ignored \citep{Laird_1987}. Recent work constructs empirical Bayes intervals for site means or posterior quantities that account for this uncertainty \citep{Armstrong_2022, Ignatiadis_2022}. In multi-site trials, \citet{Lee_2025} examine how best to estimate and rank site-specific ITT effects. None of these studies considers effects adjusted for estimated local conditions. Inference on LRE requires both this adjustment and intervals that reflect the uncertainty in the estimated site-level summaries. To our knowledge, this combination remains largely unaddressed.

\subsection{Significance and Contributions}
\label{sec:contributions}

We address both challenges using Bayesian hierarchical models that jointly estimate site-level parameters and their cross-site distribution, yielding a posterior distribution for each site's LRE. Our contributions are threefold, spanning methodology and application.

\textit{Methodologically}, (1) our models account for estimation uncertainty in site-level summaries rather than treating their sample estimates as known. This reduces bias arising from noisy regressors and correlated measurement errors in the second-stage regression. In simulations calibrated to the NJCS, the Bayesian hierarchical models fitted to individual-level data achieve the lowest root mean squared error (RMSE) for LREs in every setting except one with only categorical covariates, where they perform similarly to the best alternative. The gains are largest when within-site sample sizes are small. We also develop a Bayesian hierarchical model fitted to site-level summaries, allowing estimation when individual-level data are unavailable.

(2) Joint Bayesian inference provides posterior distributions for LREs and their cross-site standard deviation, accounting for uncertainty in the estimation of site-level parameters and their cross-site distribution. In simulations, the resulting 95\% credible intervals achieve near-nominal coverage. The posterior also provides probabilities of tier classification for each site.

\textit{For the application}, (3) we reanalyze the NJCS to estimate the LRE of each Job Corps center, accounting for uncertainty in site-level summaries. We report each center's posterior probabilities of belonging to the lower, middle, or higher LRE tier, providing information for decisions about contract renewal, oversight, and program improvement. The approach extends to multi-site randomized evaluations in the social sciences, workforce training, public education, and healthcare, supporting further investigation of which organizational practices to discontinue, maintain, or expand.

\section{The Local Relative Effectiveness (LRE): Estimand and Model}
\label{sec:lre}

Consider a multi-site randomized trial with $J$ sites, each containing $n_j$ individuals. For individual $i$ at site $j$, let $Z_{ij} \in \{0,1\}$ denote treatment assignment and $Y_{ij}$ the observed outcome. Let $\boldsymbol{X}_{ij} \in \mathbb{R}^p$ and $\boldsymbol{U}_{ij}$ denote observed and unobserved pre-treatment covariates (e.g., age, race, and prior labor market experience), respectively. Either vector may include site-level characteristics shared by all individuals within a site (e.g., urban versus rural location). Together, they characterize the local conditions we seek to account for. Let $n_{1j}$ and $n_{0j}$ denote the treatment and control sample sizes at site $j$, with total $n_{1j} + n_{0j} = n_j$. Under the potential outcomes framework and the stable unit treatment value assumption (SUTVA) \citep{Rubin_1980, Holland_1986}, let $Y_{ij}(0)$ and $Y_{ij}(1)$ denote the potential outcomes under control and treatment, respectively. The observed outcome is then $Y_{ij} = Z_{ij}Y_{ij}(1) + (1 - Z_{ij})Y_{ij}(0)$.

\subsection{The Estimand}
\label{sec:estimand}

Let $\mu_{0j} = \mathbb{E}[Y_{ij}(0)\mid j]$ and $\mu_{1j} = \mathbb{E}[Y_{ij}(1)\mid j]$ denote the mean potential outcomes at site $j$ under control and treatment, respectively, and let $\delta_j = \mu_{1j} - \mu_{0j}$ be the site-specific ITT effect. Let $F_{Xj}$ and $F_{Uj}$ denote the distributions of $\boldsymbol{X}_{ij}$ and $\boldsymbol{U}_{ij}$, respectively, within site $j$. The pair $(F_{Xj}, F_{Uj})$ summarizes the composition of site $j$ and other local conditions under which its organization operates. The site-specific covariate mean vector $\boldsymbol{\mu}_{Xj} = (\mu_{X1j}, \ldots, \mu_{Xpj})^T$ is a function of $F_{Xj}$. We take $F_{Uj}$ to be the ideal set of unobserved conditions that one would adjust for if they were measured. 

Comparing $\delta_j$ directly across sites confounds organizational effectiveness with local ecological conditions. The quantity that removes this confounding is the oracle LRE,
\begin{equation}
  \theta^{\ast}_j = \delta_j - \mathbb{E}[\delta_j \mid F_{Xj},\, F_{Uj}],
  \label{eq:lre_oracle}
\end{equation}
the deviation of the site's ITT effect from the average ITT effect among sites with the same observed and unobserved pre-treatment conditions. Because $F_{Uj}$ is unobserved, $\theta^{\ast}_j$ cannot be computed from data. Following \citet{Hong2026AJE}, we instead adjust for the site mean control outcome $\mu_{0j}$ together with $\boldsymbol{\mu}_{Xj}$, which defines the pseudo LRE,
\begin{equation}
  \theta_j = \delta_j - \mathbb{E}[\delta_j \mid \mu_{0j},\, \boldsymbol{\mu}_{Xj}].
  \label{eq:lre_def}
\end{equation}
The rationale is that $\mu_{0j}$ is the average outcome participants at site $j$ would have without the program. It is unaffected by the program and thus pre-treatment in nature, reflecting both $F_{Xj}$ and $F_{Uj}$. Conditioning on $\mu_{0j}$ therefore adjusts for unobserved conditions to the extent that they are captured by the control mean. The oracle LRE and the pseudo LRE coincide when the local conditions are related to the expected ITT effect only through $\boldsymbol{\mu}_{Xj}$ and $\mu_{0j}$, namely when
\begin{equation}
  \mathbb{E}[\delta_j \mid F_{Xj},\, F_{Uj}] = \mathbb{E}[\delta_j \mid \mu_{0j},\, \boldsymbol{\mu}_{Xj}].
  \label{eq:lre_equiv}
\end{equation}
Under Condition~\eqref{eq:lre_equiv}, after accounting for $\mu_{0j}$ and $\boldsymbol{\mu}_{Xj}$, the remaining variation in $\delta_j$ reflects organizational effectiveness. Our inference targets the pseudo LRE $\theta_j$ in Equation~\eqref{eq:lre_def} and does not rely on Condition~\eqref{eq:lre_equiv}. This condition is required only to interpret $\theta_j$ as the oracle LRE $\theta_j^{\ast}$. Hereafter, LRE refers to the pseudo LRE $\theta_j$.

\subsection{Identification and Generative Model}
\label{sec:genmodel}

Identification of $\theta_j$ has two parts. Within-site randomization identifies $\delta_j$ and $\mu_{0j}$ at each site by design, as it guarantees ignorability $Z_{ij} \independent (Y_{ij}(0), Y_{ij}(1)) \mid j$ and positivity $0 < P(Z_{ij} = 1 \mid j) < 1$. The ecological adjustment $\mathbb{E}[\delta_j \mid \mu_{0j}, \boldsymbol{\mu}_{Xj}]$ is then determined by how these site-level quantities are jointly distributed across sites. Proposition~\ref{prop:ident} states the result and the proof is provided in Appendix~\ref{app:proofs}. Apart from a finite-mean condition, it requires no assumption on the distribution of the site-level parameters or the individual-level covariates.

\begin{proposition}[Identification of the LRE]
\label{prop:ident}
Suppose that sites are independent draws from a population of sites and that, within each site, participants are independent draws from a site-specific distribution with treatment assigned at random, so that ignorability and positivity hold. Suppose also that $\mathbb{E}|\delta_j| < \infty$. Then $\delta_j$, $\mu_{0j}$, and $\boldsymbol{\mu}_{Xj}$ are functions of the distribution of the observed data $(Y_{ij}, Z_{ij}, \boldsymbol{X}_{ij})$ at site $j$, and the conditional mean $\mathbb{E}[\delta_j \mid \mu_{0j}, \boldsymbol{\mu}_{Xj}]$ is a function of the distribution of these site-level quantities across sites. Hence the LRE $\theta_j$ in Equation~\eqref{eq:lre_def} is identified.
\end{proposition}

For estimation and inference, we specify a generative model with two layers. The between-site layer describes how site-level parameters vary across sites. The within-site layer describes how individual outcomes arise given those parameters. Together they yield the joint distribution of the site-level parameters and the individual-level data, which underlies our estimation strategies. 

Collecting the site-level quantities into $\boldsymbol{\phi}_j = (\delta_j, \mu_{0j}, \boldsymbol{\mu}_{Xj}^T)^T$, we impose a multivariate normal model across sites,
\begin{equation}
  \boldsymbol{\phi}_j \overset{\mathrm{iid}}{\sim}
  \mathcal{N}_{p+2}(\boldsymbol{m},\, \boldsymbol{\Omega}),
  \quad
  \boldsymbol{m} = (\delta,\, \mu_0,\, \boldsymbol{\mu}_X^T)^T,
  \quad
  \boldsymbol{\Omega} =
  \begin{pmatrix}
    \Omega_{11} & \boldsymbol{\omega}_{21}^T \\
    \boldsymbol{\omega}_{21} & \boldsymbol{\Omega}_{22}
  \end{pmatrix},
  \quad 1 \le j \le J.
  \label{eq:phi_normal}
\end{equation}
The vector $\boldsymbol{m}$ collects the overall means $\delta = \mathbb{E}[\delta_j]$, $\mu_0 = \mathbb{E}[\mu_{0j}]$, and $\boldsymbol{\mu}_X = \mathbb{E}[\boldsymbol{\mu}_{Xj}]$. In the between-site covariance matrix $\boldsymbol{\Omega}$, $\Omega_{11} = \mathrm{Var}(\delta_j)$, $\boldsymbol{\omega}_{21}$ is the covariance vector between $\delta_j$ and $(\mu_{0j},\,\boldsymbol{\mu}_{Xj}^T)^T$, and $\boldsymbol{\Omega}_{22}$ is the covariance matrix of $(\mu_{0j},\,\boldsymbol{\mu}_{Xj}^T)^T$.

The normal model~\eqref{eq:phi_normal} is not needed for identification. Its role is to give a closed form for the conditional mean, as follows.

\begin{lem}[Gaussian closed form]
\label{lem:gauss}
Write $\boldsymbol{\phi}_{2j} = (\mu_{0j}, \boldsymbol{\mu}_{Xj}^T)^T$ and $\boldsymbol{m}_2 = (\mu_0, \boldsymbol{\mu}_X^T)^T$. Under the between-site model~\eqref{eq:phi_normal},
\begin{align}
  \mathbb{E}[\delta_j \mid \boldsymbol{\phi}_{2j}]
  &= \delta + \boldsymbol{\beta}^T(\boldsymbol{\phi}_{2j} - \boldsymbol{m}_2),
  \qquad
  \boldsymbol{\beta} := (\beta_0,\, \boldsymbol{\beta}_1^T)^T
  = \boldsymbol{\Omega}_{22}^{-1}\boldsymbol{\omega}_{21},
  \label{eq:lem_slope} \\
  \mathrm{Var}(\delta_j \mid \boldsymbol{\phi}_{2j})
  &= \sigma_\theta^2,
  \qquad
  \sigma_\theta := \bigl(\Omega_{11}
    - \boldsymbol{\omega}_{21}^T\boldsymbol{\Omega}_{22}^{-1}
      \boldsymbol{\omega}_{21}\bigr)^{1/2}.
  \label{eq:sigma_theta}
\end{align}
Consequently $\theta_j = \delta_j - \mathbb{E}[\delta_j \mid \boldsymbol{\phi}_{2j}]$ has mean zero, variance $\sigma_\theta^2$, and is uncorrelated with $\boldsymbol{\phi}_{2j}$.
\end{lem}

\noindent
By Lemma~\ref{lem:gauss}, the conditional expectation in the LRE definition~\eqref{eq:lre_def} is linear under the normal model~\eqref{eq:phi_normal}, so
\begin{equation}
  \theta_j
  = \delta_j
    - \bigl[\delta
    + \beta_0(\mu_{0j} - \mu_0)
    + \boldsymbol{\beta}_1^T(\boldsymbol{\mu}_{Xj} - \boldsymbol{\mu}_X)\bigr],
  \label{eq:theta_closed}
\end{equation}
where $\beta_0$ and $\boldsymbol{\beta}_1$ are the ecological regression slopes on the control mean and the covariate means. The standard deviation $\sigma_\theta$ measures how widely $\theta_j$ varies across sites after this adjustment.

Given the site-level parameters from the between-site model~\eqref{eq:phi_normal}, individual-level covariates and potential outcomes follow
\begin{subequations}\label{eq:genmodel}
\begin{align}
  \boldsymbol{X}_{ij} \mid \boldsymbol{\mu}_{Xj},\,\boldsymbol{\Sigma}_{Xj}
    &\overset{\mathrm{iid}}{\sim}
    \mathcal{N}_p\!\bigl(\boldsymbol{\mu}_{Xj},\;\boldsymbol{\Sigma}_{Xj}\bigr),
    \quad 1 \le i \le n_j,
    \label{eq:genmodel_X} \\
  \begin{pmatrix} Y_{ij}(0) \\ Y_{ij}(1) \end{pmatrix} \bigg| \boldsymbol{X}_{ij}
    &\overset{\mathrm{ind}}{\sim}
    \mathcal{N}_2\!\left(
      \begin{pmatrix}
        \mu_{0j} + \boldsymbol{\alpha}_{0j}^T(\boldsymbol{X}_{ij} - \boldsymbol{\mu}_{Xj}) \\
        \mu_{0j} + \delta_j + \boldsymbol{\alpha}_{1j}^T(\boldsymbol{X}_{ij} - \boldsymbol{\mu}_{Xj})
      \end{pmatrix},\;
      \boldsymbol{\Upsilon}_j
    \right),
    \label{eq:genmodel_Y}
\end{align}
\end{subequations}
where $\boldsymbol{\Sigma}_{Xj}$ is the within-site covariance matrix of covariates $\bs{X}_{ij}$; the within-site regression slopes $\boldsymbol{\alpha}_{0j}$ and $\boldsymbol{\alpha}_{1j}$ represent the linear relationships between the covariates and the potential outcomes in the control and treatment arms, respectively; and $\boldsymbol{\Upsilon}_j$ is the $2\times 2$ residual covariance matrix of the potential outcomes. The off-diagonal element of $\boldsymbol{\Upsilon}_j$ represents the covariance between $Y_{ij}(0)$ and $Y_{ij}(1)$. Because only one potential outcome is observed per individual, this covariance does not enter the observed-data likelihood and is not identifiable \citep{Holland_1986}. Since $\theta_j$ depends only on site-level means and between-site parameters, we model only the diagonal elements $\sigma_{0j}^2$ and $\sigma_{1j}^2$.

The off-diagonal element of $\boldsymbol{\Upsilon}_j$, the covariance between $Y_{ij}(0)$ and $Y_{ij}(1)$ for the same individual, is not identifiable because each individual is observed under one arm only \citep{Holland_1986}. The LRE does not involve it, so we model only the diagonal elements $\sigma_{0j}^2$ and $\sigma_{1j}^2$.

The within-site parameters $\boldsymbol{\alpha}_{zj}$, $\boldsymbol{\Upsilon}_j$, and $\boldsymbol{\Sigma}_{Xj}$ are drawn independently across sites from distributions indexed by a hyper-parameter vector $\boldsymbol{\psi}$, and are independent of $\boldsymbol{\phi}_j$ given $\boldsymbol{\psi}$. Section~\ref{sec:methods} builds the Bayesian hierarchical models that jointly estimate $\boldsymbol{m}$, $\boldsymbol{\Omega}$, and $\boldsymbol{\psi}$. The linear expression in~\eqref{eq:theta_closed} relies on normality assumptions that may not hold in practice. Section~\ref{sec:simulation} evaluates LRE recovery when they are violated.

\subsection{Interpreting and Comparing LREs}
\label{sec:interpretation}

LRE is straightforward to interpret when organizations operate under the same local conditions, that is, the same $(\mu_{0j}, \boldsymbol{\mu}_{Xj})$. Such organizations share the benchmark $\mathbb{E}[\delta_j \mid \mu_{0j}, \boldsymbol{\mu}_{Xj}]$, so differences in their LREs equal differences in their ITT effects. A higher LRE therefore indicates a larger positive program impact than organizations at other sites under the same conditions.

Comparisons across different conditions are less straightforward because each LRE is measured against a different benchmark. By construction, $\theta_j$ has mean zero at every value of $(\mu_{0j}, \boldsymbol{\mu}_{Xj})$ and is mean independent of, and hence uncorrelated with, these conditions. Thus, the LRE captures variation in organizational effectiveness unrelated to local conditions, while variation in organizational practice that is systematically associated with them is absorbed into the benchmark. This parallels an issue raised by \citet{raudenbush1995estimation} for school value-added measures: effects of organizational practice correlated with covariates are partly attributed to those covariates. For example, suppose organizations under one set of conditions generally employ more effective practices and provide higher-quality services than those under another. An organization with a positive LRE in the second group could rank above one with a negative LRE in the first, even if the latter's practices might be more effective in absolute terms. This does not indicate bias in identifying $\theta_j$, which measures effectiveness relative to organizations under similar conditions, but it limits comparisons across conditions. Such comparisons generally require assumptions about how organizations would perform under conditions they do not face, which the observed data may not support.

Three versions of LRE can be considered when comparing organizations: the original LRE, $\theta_j$; the standardized LRE, which divides $\theta_j$ by its standard deviation among sites under the same conditions; and the quantile-transformed LRE, which is the percentile rank of $\theta_j$ among those sites. Under the between-site normal model~\eqref{eq:phi_normal}, $\theta_j$ is independent of the local conditions $\boldsymbol{\phi}_{2j} = (\mu_{0j}, \boldsymbol{\mu}_{Xj})$, so it follows the same $\mathcal{N}(0,\sigma_\theta^2)$ distribution regardless of $\boldsymbol{\phi}_{2j}$. Thus, all three measures yield identical site rankings. More generally, the rankings may differ if the LRE distribution varies with the conditions, for example, in its variance. We use the original LRE throughout, including for the tier probabilities in Sections~\ref{sec:simulation} and~\ref{sec:application}, which should be interpreted with this limitation in mind.

\section{Our Proposed Methods: Bayesian Hierarchical Models}
\label{sec:methods}

The generative model treats site-level parameters as random draws from a population distribution. Our Bayesian hierarchical models jointly estimate all unknown quantities, addressing both challenges in Section~\ref{sec:challenges}: measurement error in site-level summaries and uncertainty in LRE inference. Rather than fixing $\mu_{0j}$ and $\boldsymbol{\mu}_{Xj}$ at their sample estimates, the models account for their uncertainty through the within-site likelihood for individual-level data or the sampling covariance for site-level summaries. Joint estimation propagates this uncertainty into the posterior distributions of $\theta_j$ and $\sigma_\theta$, yielding credible intervals and tier probabilities that account for uncertainty in local conditions. Section~\ref{sec:eiv} examines how joint estimation addresses plug-in bias.

Building on this, we construct three Bayesian hierarchical models (BHMs): two fitted to individual-level data and one, Site-Sum-Bayes, fitted to site-level summaries. The two individual-data BHMs differ in two assumptions: whether the within-site predictor-outcome slopes and residual variances are constant across sites, and whether the within-site covariate covariance matrix is shared across sites. The model names reflect these choices: the prefix Y refers to the potential outcome model in~\eqref{eq:genmodel_Y}, with YV indicating site-varying slopes and residual variances and YC indicating constant ones; the prefix X refers to the covariate model in~\eqref{eq:genmodel_X}, with XHet indicating site-specific covariance matrices and XHom indicating a shared covariance matrix. Here, we focus on a contrast between two models: YV-XHet, which allows all these quantities to vary across sites, and YC-XHom, which holds them constant. The two intermediate models, YV-XHom and YC-XHet, are described in Appendix~\ref{app:other_models}. Throughout, $\sigma$, $\tau$, and $s$ denote within-site noise, between-site deviations, and fixed hyper-prior scales, respectively. All priors are weakly informative and calibrated to the application scale, with justifications provided in Appendix~\ref{app:priors}.

\subsection{Bayesian Hierarchical Models with Individual-Level Data}
\label{sec:indiv_models}

\begin{itemize}
\item \textbf{YV-XHet} (site-varying predictor-outcome slopes and residual variances, heterogeneous covariate covariance matrices) is the most flexible specification. It has the same structure as the generative model of Section~\ref{sec:genmodel} and adds hierarchical priors on its parameters. 

\textit{Between-site hyper-priors.} The between-site mean vector $\boldsymbol{m}$ and covariance matrix $\boldsymbol{\Omega}$ are assigned the hyper-priors
\begin{subequations}\label{eq:v2sx}
\begin{align}
  \boldsymbol{m}
    &\sim \mathcal{N}_{p+2}(\boldsymbol{m}_0,\;\mathrm{diag}(\boldsymbol{s}_m^2)),
    \label{eq:prior_m} \\
  \boldsymbol{\Omega}
    &= \mathrm{diag}(\boldsymbol{\tau})\,\boldsymbol{R}\,\mathrm{diag}(\boldsymbol{\tau}),
    \quad
    \tau_k \overset{\mathrm{ind}}{\sim} \mathcal{N}_+(0,\,s_k^2),
    \quad
    \boldsymbol{R} \sim \mathrm{LKJ}(2),
    \quad 1 \le k \le p+2,
    \label{eq:prior_Omega}
\end{align}
where $\boldsymbol{\tau} = (\tau_1, \ldots, \tau_{p+2})^T$ collects the between-site standard deviations, $\boldsymbol{R}$ is the between-site correlation matrix, and $\mathcal{N}_+(0, s^2)$ is the half-normal distribution with scale $s$. The $\mathrm{LKJ}(2)$ prior mildly regularizes $\boldsymbol{R}$ toward the identity \citep{Lewandowski_2009}, and the half-normal priors follow \citet{Gelman_2006}. Appendix~\ref{app:priors} reports all hyper-parameter values.

\textit{Within-site slopes and residual variances.} The site-specific slopes $\boldsymbol{\alpha}_{zj}$ and log residual variances $\log\sigma_{zj}^2$ are drawn from hierarchical distributions that pool information across sites. For $1 \le j \le J$ and $z \in \{0,1\}$,
\begin{align}
  \boldsymbol{\alpha}_{zj}
    &\overset{\mathrm{iid}}{\sim}
    \mathcal{N}_p(\bar{\boldsymbol{\alpha}}_z,\, \tau_\alpha^2 I_p),
    \quad
    \bar{\boldsymbol{\alpha}}_z \sim \mathcal{N}_p(\boldsymbol{0},\,s_\alpha^2 I_p),
    \quad
    \tau_\alpha \sim \mathcal{N}_+(0,\,s_{\tau_\alpha}^2),
    \label{eq:alpha_prior} \\
  \log\sigma_{zj}^2
    &\overset{\mathrm{iid}}{\sim}
    \mathcal{N}(\mu_{\log\sigma,z},\tau_{\log\sigma}^2),
    \;\;
    \mu_{\log\sigma,z} \sim \mathcal{N}(m_{\log\sigma,z},s_{\mu_{\log\sigma}}^2),
    \;\;
    \tau_{\log\sigma} \sim \mathcal{N}_+(0,s_{\tau_{\log\sigma}}^2),
    \label{eq:sigma_prior}
\end{align}
where $\bar{\boldsymbol{\alpha}}_z$ is the mean slope vector for arm $z$, $\tau_\alpha$ is the between-site standard deviation of the slopes, and $I_p$ is the $p\times p$ identity matrix. The scalar $\tau_\alpha$ is common to all covariates and both arms, which presumes covariates on comparable scales.
Similarly, $\mu_{\log\sigma,z}$ and $\tau_{\log\sigma}$ are the between-site mean and standard deviation of the log residual variance, and $m_{\log\sigma,z}$ is a fixed prior center. The $\sigma_{zj}^2$ are the diagonal elements of $\boldsymbol{\Upsilon}_j$, whose off-diagonal element is not modeled (Section~\ref{sec:genmodel}). Modeling the variances on the log scale keeps them positive.

\textit{Within-site covariate covariance matrix.} The site-specific covariate covariance matrix $\boldsymbol{\Sigma}_{Xj}$ is separated into standard deviations and correlations. For $1 \le j \le J$ and $1 \le k \le p$,
\begin{align}
  \boldsymbol{\Sigma}_{Xj}
    &= \mathrm{diag}(\boldsymbol{\sigma}_{Xj})\,\boldsymbol{R}_{Xj}\,
      \mathrm{diag}(\boldsymbol{\sigma}_{Xj}),
    \quad
    \boldsymbol{R}_{Xj} \overset{\mathrm{iid}}{\sim} \mathrm{LKJ}(2),
    \label{eq:sigmax_prior} \\
  \sigma_{Xjk}
    &\overset{\mathrm{ind}}{\sim} \mathrm{LogNormal}(\mu_{\log\sigma_X, k},\,
      \tau_{\log\sigma_X}^2),
    \quad
    \boldsymbol{\mu}_{\log\sigma_X} \sim \mathcal{N}_p(\boldsymbol{0},\, s_{\mu_X}^2 I_p).
    \label{eq:sigmax_hyper}
\end{align}
\end{subequations}
The vector $\boldsymbol{\sigma}_{Xj} = (\sigma_{Xj1}, \ldots, \sigma_{Xjp})^T$ collects the within-site covariate standard deviations, with $\boldsymbol{\mu}_{\log\sigma_X} = (\mu_{\log\sigma_X, 1}, \ldots, \mu_{\log\sigma_X, p})^T$ and $\tau_{\log\sigma_X} \sim \mathcal{N}_+(0,\, s_{\tau_X}^2)$. The within-site correlation matrix $\boldsymbol{R}_{Xj}$ is distinct from the between-site correlation matrix $\boldsymbol{R}$ in~\eqref{eq:prior_Omega}.

\item \textbf{YC-XHom} (constant predictor-outcome slopes and residual variances, homogeneous covariate covariance matrix) is the simplest specification. The model keeps the between-site hyper-priors~\eqref{eq:prior_m}--\eqref{eq:prior_Omega} of YV-XHet, so the two models treat the site-level parameters $\boldsymbol{\phi}_j$ in the same way. They differ only in the within-site layer.

\textit{Constant slopes and residual variances.} The slopes and residual standard deviations are constrained to be constant across sites, $\boldsymbol{\alpha}_{zj} = \boldsymbol{\alpha}_z$ and $\sigma_{zj} = \sigma_z$ for all $j$, with priors
\begin{equation}
  \boldsymbol{\alpha}_z \sim \mathcal{N}_p(\boldsymbol{0},\,s_\alpha^2 I_p),
    \qquad
  \sigma_z \sim \mathcal{N}_+(0,\,s_\sigma^2),
    \qquad z \in \{0,1\}.
    \label{eq:v1_priors}
\end{equation}
These priors replace the hierarchical priors~\eqref{eq:alpha_prior} and~\eqref{eq:sigma_prior} of YV-XHet.

\textit{Shared covariate covariance matrix.} The covariate covariance matrix is shared across sites, $\boldsymbol{\Sigma}_{Xj} = \boldsymbol{\Sigma}_X$ for all $j$. For $1 \le k \le p$, the prior is
\begin{equation}
  \boldsymbol{\Sigma}_X = \mathrm{diag}(\boldsymbol{\sigma}_X)\,
    \boldsymbol{R}_X\,\mathrm{diag}(\boldsymbol{\sigma}_X),
    \quad
  \boldsymbol{R}_X \sim \mathrm{LKJ}(2),
    \quad
  \sigma_{Xk} \overset{\mathrm{iid}}{\sim} \mathcal{N}_+(0,\,s_{\sigma_X}^2),
    \label{eq:sigmax_shared}
\end{equation}
where $\boldsymbol{\sigma}_X = (\sigma_{X1}, \ldots, \sigma_{Xp})^T$ collects the common within-site covariate standard deviations and $\boldsymbol{R}_X$ is the common within-site correlation matrix. This prior replaces the site-specific prior~\eqref{eq:sigmax_prior}--\eqref{eq:sigmax_hyper} of YV-XHet. YC-XHom matches the data when sites share the same within-site structure. When that structure varies across sites, its within-site layer is misspecified, and our simulation examines how much this matters for the LRE.
\end{itemize}

\subsection{Site-Level Bayesian Model with Summary Statistics}
\label{sec:site_model}

In many applied settings, access to individual-level records is restricted by privacy regulations or data-sharing agreements. The site-level model, which we call Site-Sum-Bayes, is designed for this setting and uses only the site-level summary statistics $\hat{\boldsymbol{\phi}}_j$. The site-level approach of \citet{Lee_2025} estimates site-specific effects from summary statistics but does not adjust for local ecological conditions. Site-Sum-Bayes adds this adjustment, treating the control mean and covariate means as estimated inputs and carrying their uncertainty into the estimation and inference of $\theta_j$.

Site-Sum-Bayes uses the summary vector $\hat{\boldsymbol{\phi}}_j=(\bar{Y}_{1j}-\bar{Y}_{0j},\,\bar{Y}_{0j},\,\bar{\boldsymbol{X}}_j^T)^T$, whose components are unbiased for $\boldsymbol{\phi}_j=(\delta_j,\mu_{0j},\boldsymbol{\mu}_{Xj}^T)^T$ under within-site randomization and the covariate model~\eqref{eq:genmodel_X}. For large $n_j$, the central limit theorem motivates
\begin{equation}
  \hat{\boldsymbol{\phi}}_j\mid\boldsymbol{\phi}_j,\hat{\boldsymbol{\Sigma}}_{\phi j}
  \stackrel{\text{approx.}}{\sim}
  \mathcal{N}_{p+2}(\boldsymbol{\phi}_j,\hat{\boldsymbol{\Sigma}}_{\phi j}),
  \quad 1 \le j \le J,
  \label{eq:phihat_approx}
\end{equation}
where $\hat{\boldsymbol{\Sigma}}_{\phi j}$ is estimated from within-site data and treated as known. This approximation does not require individual-level normality.

Let $\boldsymbol{g}_j=(\bar{Y}_{1j},\bar{Y}_{0j},\bar{\boldsymbol{X}}_j^T)^T$, where $\bar{\boldsymbol{X}}_j=n_j^{-1}\sum_{i=1}^{n_j}\boldsymbol{X}_{ij}$. The summary and its estimated sampling covariance are
\begin{equation}
  \hat{\boldsymbol{\phi}}_j=\boldsymbol{\Xi}\boldsymbol{g}_j,
  \qquad
  \hat{\boldsymbol{\Sigma}}_{\phi j}
  =\boldsymbol{\Xi}\hat{\boldsymbol{G}}_j\boldsymbol{\Xi}^T,
  \qquad
  \boldsymbol{\Xi}=
  \begin{pmatrix}
    1 & -1 & \boldsymbol{0}_{1\times p}\\
    0 & 1 & \boldsymbol{0}_{1\times p}\\
    \boldsymbol{0}_{p\times1} & \boldsymbol{0}_{p\times1} & I_p
  \end{pmatrix},
  \label{eq:Sigma_j}
\end{equation}
where $\hat{\boldsymbol{G}}_j$ estimates the sampling covariance of $\boldsymbol{g}_j$:
\begin{equation}
  \hat{\boldsymbol{G}}_j=
  \begin{pmatrix}
    \hat{v}_{1j}/n_{1j} & 0 & \hat{\boldsymbol{s}}_{YX,1j}^T/n_j\\
    0 & \hat{v}_{0j}/n_{0j} & \hat{\boldsymbol{s}}_{YX,0j}^T/n_j\\
    \hat{\boldsymbol{s}}_{YX,1j}/n_j & \hat{\boldsymbol{s}}_{YX,0j}/n_j &
    \hat{\boldsymbol{\Sigma}}_{Xj}/n_j
  \end{pmatrix}.
  \label{eq:Gj_matrix}
\end{equation}
Here $\hat{v}_{zj}$ and $\hat{\boldsymbol{s}}_{YX,zj}$ denote the sample variance of $Y$ and the sample covariance between $Y$ and $\boldsymbol{X}$ in arm $z$, respectively, while $\hat{\boldsymbol{\Sigma}}_{Xj}$ is the sample covariance matrix of $\boldsymbol{X}$ pooled across both arms at site $j$. The covariance between $\bar{Y}_{1j}$ and $\bar{Y}_{0j}$ is zero because they are computed from disjoint samples. The cross-covariances with $\bar{\boldsymbol{X}}_j$ are nonzero because $\bar{\boldsymbol{X}}_j$ includes participants from both arms. With the standard $n-1$ correction, each sample variance and covariance is unbiased, making $\hat{\boldsymbol{\Sigma}}_{\phi j}$ an unbiased estimator of the sampling covariance of $\hat{\boldsymbol{\phi}}_j$. Appendix~\ref{app:site_cov} provides the derivation.

\subsection{Addressing Plug-in Bias}
\label{sec:eiv}

In the site-level model, a least-squares procedure 
simply regresses estimated site-specific ITT effects $\hat{\delta}_j$ on estimated local conditions $\hat{\boldsymbol{\phi}}_{2j} = (\bar{Y}_{0j}, \bar{\boldsymbol{X}}_j^T)^T$. Because this regression treats noisy site-level summaries as error-free, the slopes are subject to finite sample bias, which we refer to as plug-in bias. Below we characterize this bias in the OLS estimation and explain how the Bayesian models account for it.

Write $\hat{\delta}_j = \delta_j + e_{\delta j}$ and $\hat{\boldsymbol{\phi}}_{2j} = \boldsymbol{\phi}_{2j} + \boldsymbol{e}_{\mu j}$, where the sampling errors have conditional mean zero given the site-level parameters and are independent across sites. 
Let $\boldsymbol{\Sigma}_{\phi j}$, the population counterpart of $\hat{\boldsymbol{\Sigma}}_{\phi j}$, denote their joint sampling covariance, with components $\boldsymbol{\Sigma}_{\mu\mu,j} = \mathrm{Var}(\boldsymbol{e}_{\mu j})$ and $\boldsymbol{\sigma}_{\delta\mu,j} = \mathrm{Cov}(\boldsymbol{e}_{\mu j},e_{\delta j})$. 
Under the linear model in Lemma~\ref{lem:gauss}, as $J \to \infty$ with site sizes fixed, the least-squares slope $\boldsymbol{b}$ from regressing $\hat{\delta}_j$ on $\hat{\boldsymbol{\phi}}_{2j}$ converges to
\begin{equation}
  \operatorname*{plim} \boldsymbol{b}
  = (\boldsymbol{\Omega}_{22} + \bar{\boldsymbol{\Sigma}}_{\mu\mu})^{-1}
    (\boldsymbol{\omega}_{21} + \bar{\boldsymbol{\sigma}}_{\delta\mu}),
  \label{eq:plim_b}
\end{equation}
where $\bar{\boldsymbol{\Sigma}}_{\mu\mu} = \lim_{J \to \infty} J^{-1}\sum_{j=1}^{J}\boldsymbol{\Sigma}_{\mu\mu,j}$ and $\bar{\boldsymbol{\sigma}}_{\delta\mu} = \lim_{J \to \infty} J^{-1}\sum_{j=1}^{J}\boldsymbol{\sigma}_{\delta\mu,j}$ are assumed to exist (Appendix~\ref{app:proofs}). The true ecological slope is $\boldsymbol{\beta} = \boldsymbol{\Omega}_{22}^{-1}\boldsymbol{\omega}_{21}$. Thus, sampling error affects both the regressor covariance and its covariance with the outcome, generally making $\boldsymbol{b}$ inconsistent for $\boldsymbol{\beta}$. The plug-in LRE, $\hat{\theta}_j^{\,\text{plug-in}} = \hat{\delta}_j - \hat{\delta} - \boldsymbol{b}^T(\hat{\boldsymbol{\phi}}_{2j} - \hat{\boldsymbol{m}}_2)$, where $\hat{\delta}$ and $\hat{\boldsymbol{m}}_2$ are sample estimates, consequently uses a biased adjustment for local conditions. Both sampling covariance terms are of order $n_j^{-1}$, so the resulting slope bias is larger for smaller sites and generally persists as the number of sites increases while site sizes remain fixed.

The 2S-ME procedure of \citet{Hong2026AJE} replaces $\bar{Y}_{0j}$ with its empirical Bayes estimate from Step 1, so the limit in~\eqref{eq:plim_b} does not apply to it directly. Shrinkage lowers the sampling variance of the control-mean predictor and thus reduces attenuation. It does not remove the bias. The error in the shrunken control mean remains negatively correlated with the error in $\hat{\delta}_j$, and the covariate means $\bar{\boldsymbol{X}}_j$ enter the second-stage regression without shrinkage. 

In contrast, the Bayesian models take a principled approach to accounting for the measurement error. We treat $\boldsymbol{\phi}_{2j}$ as unknown rather than fixing it at its sample estimate. In the site-level model, the measurement model~\eqref{eq:phihat_approx} describes the observed summaries $\hat{\boldsymbol{\phi}}_j$ conditional on the unobserved site parameters $\boldsymbol{\phi}_j$, which follow the between-site model~\eqref{eq:phi_normal}. Integrating out $\boldsymbol{\phi}_j$ gives
\begin{equation}
  \hat{\boldsymbol{\phi}}_j \sim
  \mathcal{N}_{p+2}(\boldsymbol{m},\;
  \boldsymbol{\Omega} + \boldsymbol{\Sigma}_{\phi j}).
  \label{eq:marginal_phihat}
\end{equation}
The observed covariance therefore combines between-site variation with within-site sampling error. When $\boldsymbol{\Sigma}_{\phi j}$ is known, the likelihood accounts for this sampling component when estimating $\boldsymbol{\Omega}$, avoiding the plug-in bias in~\eqref{eq:plim_b} under the assumed model. The individual-data BHMs account for the same uncertainty through their within-site likelihoods. In practice, we replace $\boldsymbol{\Sigma}_{\phi j}$ with its unbiased estimator $\hat{\boldsymbol{\Sigma}}_{\phi j}$ from Appendix~\ref{app:site_cov}, which accounts for the leading sampling-error contribution.

\section{Simulation Study}
\label{sec:simulation}
\subsection{Data-Generating Processes (DGPs)}
\label{sec:sim_design}

We generate data for $J$ sites with $n_j$ individuals per site. For each site, $\boldsymbol{\phi}_j$ is drawn from the between-site model~\eqref{eq:phi_normal}, with $\boldsymbol{m}$ and $\boldsymbol{\Omega}$ calibrated to the NJCS \citep{Schochet_2008}. We consider four data-generating processes (DGPs) that differ in their within-site structure (Table~\ref{tab:dgp}). The Constant and Varying DGPs follow the generative model in Section~\ref{sec:genmodel}, whereas the Categorical-$\boldsymbol{X}$ and Mixed-$\boldsymbol{X}$ DGPs violate its normality assumption for covariates. 

The \textit{Constant} DGP includes two continuous covariates ($p=2$) drawn from a bivariate normal distribution centered at the site means. Within-site slopes, residual variances, and covariate covariance matrices are constant across sites.

The \textit{Varying} DGP retains the two continuous covariates but lets the within-site structure vary. Slopes and residual variances follow hierarchical distributions, and each site has its own covariate covariance matrix.

The \textit{Categorical-$\boldsymbol{X}$} DGP tests robustness to violations of the normal covariate model~\eqref{eq:genmodel_X}. Both covariates are three-category variables representing age and baseline earnings, generated by thresholding latent normal variables with site-specific means, so category composition varies across sites. Each covariate enters through two dummy indicators, which the Bayesian models treat as normally distributed. The site-level dummy means are one-to-one functions of the latent means, so conditioning on either gives the same true $\theta_j$. However, the conditional mean of $\delta_j$ is linear in the latent means and nonlinear in the dummy means, which the models approximate linearly.

The \textit{Mixed-$\boldsymbol{X}$} DGP combines the two continuous covariates of the Constant DGP with one three-category covariate generated in the same way. Within sites, its latent variable is correlated with both continuous covariates, so the observed covariates are not jointly normal.

We next assess model specification under each DGP, considering the likelihoods separately from the priors. Under the Constant DGP, both individual-data BHMs are correctly specified: YC-XHom matches this DGP, while YV-XHet includes it as a special case. Under the Varying DGP, YV-XHet is correctly specified, whereas YC-XHom incorrectly assumes common slopes, residual variances, and covariate covariance matrices across sites. Because Site-Sum-Bayes models site summaries rather than the within-site structure, its sampling model remains approximately valid under both DGPs. Under Categorical-$\boldsymbol{X}$ and Mixed-$\boldsymbol{X}$ DGPs, however, all three Bayesian models are misspecified. With the between-site parameters $(\boldsymbol{m},\boldsymbol{\Omega})$ held fixed across these DGPs, performance differences reflect changes in the within-site outcome and covariate distributions.

For each DGP, we cross the number of sites $J\in\{100,1{,}000\}$ with the within-site sample size $n_j\in\{100,1{,}000\}$, yielding 16 design cells in total, each replicated $100$ times. Under the chosen $\boldsymbol{\Omega}$, $\mu_{0j}$ provides information about $\delta_j$ beyond $\boldsymbol{\mu}_{Xj}$, making adjustment for the control mean relevant to estimating $\theta_j$. Full specifications and parameter values are provided in Appendices~\ref{app:dgp} and~\ref{app:dgp_models}. We further examine an intermediate Varying-Slope DGP in Appendix~\ref{app:varying}.

\begin{table}[!htbp]
\caption{True Data-Generating Processes (DGPs) and Matching Analytic Models}
\label{tab:dgp}
\resizebox{\linewidth}{!}{%
\begin{tabular}{lllll}
 & Constant & Varying & Categorical-$\boldsymbol{X}$ & Mixed-$\boldsymbol{X}$ \\
\midrule
Covariates ($\boldsymbol{X}_{ij}$)
  & $\mathcal{N}_2$
  & $\mathcal{N}_2$
  & two 3-category (4 dummies)
  & $\mathcal{N}_2$ + 3-category (2 dummies) \\
Between-site vector ($\boldsymbol{\phi}_j$)
  & $\mathcal{N}_4(\boldsymbol{m}, \boldsymbol{\Omega})$
  & $\mathcal{N}_4(\boldsymbol{m}, \boldsymbol{\Omega})$
  & $\mathcal{N}_4(\boldsymbol{m}, \boldsymbol{\Omega})$
  & $\mathcal{N}_5(\boldsymbol{m}, \boldsymbol{\Omega})$ \\
Within-site slopes
  & constant
  & site-varying
  & constant
  & constant \\
Residual variances
  & constant
  & site-varying
  & constant
  & constant \\
Covariate covariance ($\boldsymbol{\Sigma}_{Xj}$)
  & constant
  & site-varying
  & --
  & constant \\
Between-site composition
  & --
  & --
  & site-varying
  & site-varying \\
\midrule
Analytic model matching the DGP & YC-XHom & YV-XHet & none & none \\
\bottomrule
\end{tabular}%
}

\vspace{2pt}
\begin{minipage}{\linewidth}
\small
\textit{Note.} The four columns correspond to the four different DGPs. A model matches a DGP when it has the same within-site structure. Dashes mark entries that do not apply. Parameter values are in Appendix~\ref{app:dgp}.
\end{minipage}
\end{table}

\subsection{Alternative Methods for Comparison}
\label{sec:comparison}

We compare the three Bayesian hierarchical models with three alternative methods that take the estimated site-level quantities as given. Two use individual-level records and one uses site-level summaries alone. Each estimates the LRE as the site-level residual, or the random deviation of the site-specific ITT effect from its predicted value given the estimated site conditions. 

\textit{Plug-in OLS} uses individual-level data in two steps. In Step 1, a within-site regression of $Y_{ij}$ on $Z_{ij}$ and $\boldsymbol{X}_{ij}$ produces a covariate-adjusted treatment effect $\hat{\delta}_j$ for each site. In Step 2, $\hat{\delta}_j$ is regressed on $\bar{Y}_{0j}$ and $\bar{\boldsymbol{X}}_j$ across sites, with the residual taken as the LRE estimate. 

\textit{2S-ME} \citep{Hong2026AJE} also proceeds in two steps using individual-level data. In Step 1, a random-intercept model fitted to control participants with adjustment for $\bar{\boldsymbol{X}}_j$ produces empirical Bayes estimates of the adjusted site-specific control mean outcome \citep{Morris_1983}. In Step 2, a mixed-effects model fitted to both treatment and control participants allows the treatment effect to vary across sites as a function of this estimated site-specific control mean outcome and $\bar{\boldsymbol{X}}_j$, with the site-specific random slope for $Z_{ij}$ as the LRE estimate. 

\textit{Site-Sum-OLS}, the least-squares counterpart of Site-Sum-Bayes, uses only site-level summaries. It estimates each site's treatment effect by the unadjusted mean difference $\bar{Y}_{1j}-\bar{Y}_{0j}$ and then applies the same site-level regression as Plug-in OLS. Appendix~\ref{app:comparison_methods} provides the estimating equations and \textsf{R} implementations, along with two post-estimation adjustments to 2S-ME: constrained Bayes (2S-ME-CB) and triple-goal (2S-ME-TG). Their results appear in Appendix~\ref{app:extra_results}.

We hypothesize that the individual-data BHMs will achieve the lowest RMSE for LREs by treating $\mu_{0j}$ and $\boldsymbol{\mu}_{Xj}$ as unknown and accounting for the plug-in bias discussed in Section~\ref{sec:eiv}. Site-Sum-Bayes is expected to be less accurate because it does not use individual-level covariate information. Among the comparators, 2S-ME should outperform Plug-in OLS because its first-step shrinkage partially reduces plug-in bias, while Site-Sum-OLS is expected to perform worst. For interval coverage, 2S-ME is expected to under-cover because its intervals (Section~\ref{sec:setup_eval}) do not account for uncertainty in $\bar{\boldsymbol{X}}_j$ \citep{Laird_1987}. In contrast, the Bayesian models incorporate this uncertainty and are expected to achieve nominal coverage.

\subsection{Implementation and Evaluation}
\label{sec:setup_eval}

All analyses use \textsf{R}. Bayesian models are fitted in Stan \citep{Carpenter_2017} via \texttt{cmdstanr}, using within-site sufficient statistics, and the 2S-ME comparator uses \texttt{lme4} \citep{Bates_2015}. Sampler settings and convergence diagnostics are reported in Appendix~\ref{app:suffstat}. Code is available at \texttt{[blinded for review]}.

We evaluate LRE estimation accuracy and interval coverage. Let $\hat{\theta}_j^{(d)}$ and $\theta_j^{(d)}$ denote the estimated and true LRE for site $j$ in replication $d=1,\ldots,D$. The root mean squared error in replication $d$ is $\mathrm{RMSE}^{(d)}=\{J^{-1}\sum_{j=1}^J(\hat{\theta}_j^{(d)}-\theta_j^{(d)})^2\}^{1/2}$, and Mean-RMSE is its average over the $D$ replications. Relative-RMSE divides each method's Mean-RMSE by that of Plug-in OLS in the same design cell, so values below one indicate better accuracy than Plug-in OLS. Pearson and Spearman correlations assess linear and rank agreement (Appendix~\ref{app:extra_results}). Cov95 is the proportion of site-specific 95\% intervals containing the true LRE, averaged across replications. For Bayesian models, we use equal-tailed credible intervals, bounded by the 2.5th and 97.5th percentiles of the posterior draws of $\theta_j$.
For 2S-ME, we use normal-approximation intervals, equal to the point estimate plus or minus 1.96 times the conditional standard deviation of the Step 2 random treatment effect \citep{Bates_2015}. This standard deviation is conditional on the estimated fixed effects and variance components.
Cov95 is not reported for the OLS comparators, which provide only point estimates.

We additionally evaluate classification of organizations into lower, middle, and upper LRE tiers. Within each replication, let $q_{0.30}$ and $q_{0.70}$ denote the 30th and 70th percentiles of the true LREs $\boldsymbol{\theta}=(\theta_1,\ldots,\theta_J)$ across the $J$ sites.
The true tier $t_j$ of the organization at site $j$ is lower if $\theta_j<q_{0.30}(\boldsymbol{\theta})$, upper if $\theta_j>q_{0.70}(\boldsymbol{\theta})$, and middle otherwise. Thus, the lower, middle, and upper tiers contain the bottom 30\%, middle 40\%, and top 30\% of organizations by true LRE. 
For the Bayesian estimators of LREs, tier probabilities are obtained by applying this rule to each posterior draw: $\hat{\pi}_j^{\mathrm{lower}}=P\{\theta_j<q_{0.30}(\boldsymbol{\theta})\mid\text{data}\}$ and $\hat{\pi}_j^{\mathrm{upper}}=P\{\theta_j>q_{0.70}(\boldsymbol{\theta})\mid\text{data}\}$, with the middle-tier probability given by the remainder $\hat{\pi}_j^{\mathrm{mid}} = 1 - \hat{\pi}_j^{\mathrm{lower}} - \hat{\pi}_j^{\mathrm{upper}}$. For example, $\hat{\pi}_j^{\mathrm{lower}}$ is the share of the posterior draws in which site $j$ falls below that draw's 30th percentile. The estimated tier $\hat{t}_j$ is the tier with the highest posterior probability.
For the comparison methods, estimated tiers are assigned by applying the same 30th and 70th percentile cutoffs to the point estimates of LREs. 
Because each LRE is defined under its own local conditions, these tiers compare deviations from potentially different benchmarks (Section~\ref{sec:interpretation}).

We assess tier recovery using three error rates, each averaged across 100 replications. 
The severe classification error rate (SCE) is the proportion of all organizations classified into the opposite tail, that is, lower-tier organizations misclassified as upper-tier or vice versa. 
The false-allocation rate for the lower tier ($\mathrm{FA}^{\mathrm{lower}}$) is the proportions of true middle- and upper-tier organizations misclassified as lower-tier.
The false-allocation rate for the upper tier ($\mathrm{FA}^{\mathrm{upper}}$) is the proportion of true lower- and middle-tier organizations misclassified as upper-tier. Lower values indicate better tier recovery. 
In the real-data analysis, we also report posterior tier probabilities to convey classification uncertainty. 

\subsection{Results}
\label{sec:results}

\begin{figure}[!htbp]
    \centering
    \includegraphics[width=\linewidth]{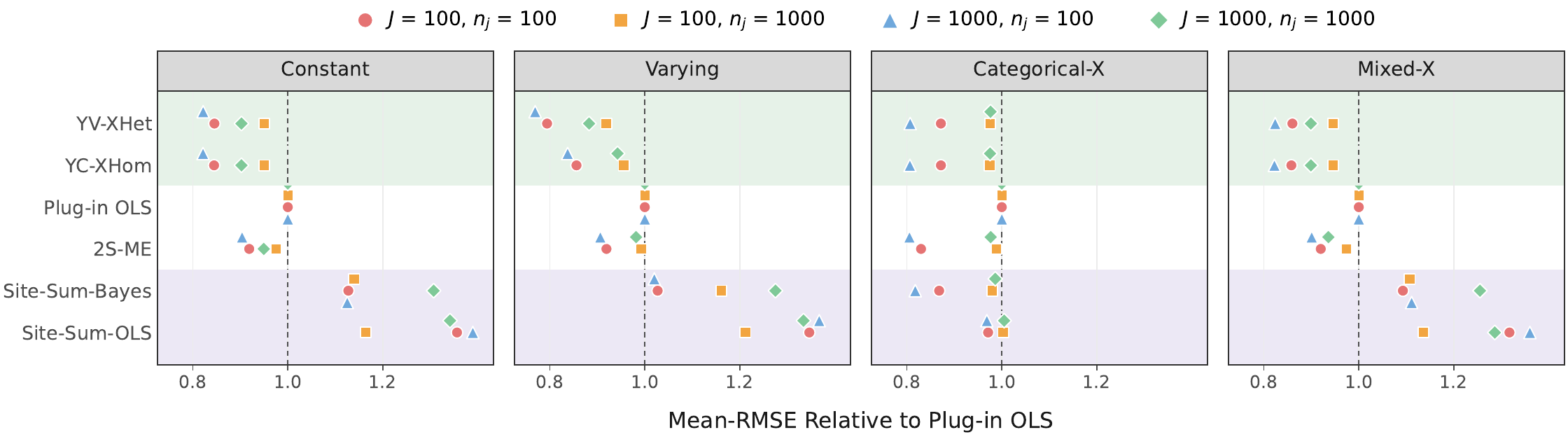}
    \caption{Mean RMSE of $\hat{\theta}_j$ relative to Plug-in OLS. Values below one (dashed line) indicate lower error than Plug-in OLS. Colors and shapes distinguish the four design cells, defined by the number of sites $(J)$ and the sample size per site $(n_j)$. Vertical offsets only separate overlapping markers. Green and purple bands group the individual-data BHMs and site-level (summary-statistic) methods, respectively.}
    \label{fig:rmse_rel}
\end{figure}

Figure~\ref{fig:rmse_rel} shows the RMSE of each estimator relative to Plug-in OLS. The two individual-data BHMs (YV-XHet and YC-XHom) outperform Plug-in OLS in every design cell. Their RMSE ratios range from about 0.77 to 0.87 at $n_j=100$ and are closer to one at $n_j=1{,}000$. Increasing $J$ from 100 to 1{,}000 has little effect. As discussed in Section~\ref{sec:eiv}, plug-in bias arises from sampling variability in site-level summaries, which is of order $n_j^{-1}$ and therefore decreases with site size rather than with the number of sites. The Bayesian models account for uncertainty in these site-level summaries. In contrast, 2S-ME corrects the control mean for measurement error by shrinkage but treats covariate means as fixed. So it generally falls between the Bayesian models and Plug-in OLS. Under the Constant, Varying, and Mixed-$\boldsymbol{X}$ DGPs, both site-Sum-Bayes and Site-Sum-OLS methods perform worse than Plug-in OLS because they omit within-site covariate adjustment, which matters when covariates explain a large share of the outcome variation. By reducing plug-in bias, Site-Sum-Bayes consistently outperforms Site-Sum-OLS, although both take the same site-level summaries as inputs.

Each individual-data BHM outperforms the competing methods in every design cell in which it is correctly specified, and in some misspecified cases as well. YC-XHom is correctly specified under the Constant DGP. YV-XHet is correctly specified under both the Constant and Varying DGPs, because it allows the slopes, residual variances, and covariate covariance matrices to vary across sites even though they do not vary under the Constant DGP. Under the Varying DGP, however, YC-XHom incorrectly constrain these quantities to be constant across sites. Accordingly, the two models perform nearly identically under the Constant DGP, whereas YV-XHet has a clearer advantage under the Varying DGP. Even under this misspecification, YC-XHom still outperforms every other competing method. Similarly, under the Mixed-X DGP, both YV-XHet and YC-XHom misspecify the within-site distribution of covariates, yet both still outperform the competing methods in every design cell.

The Categorical-$\boldsymbol{X}$ DGP is an exception to this pattern. At $n_j=100$, the Bayesian hierarchical models and 2S-ME have similar accuracy, while the two OLS comparators perform worse. At $n_j=1{,}000$, all six methods are within a few percent of one another. Unlike the Constant DGP, where covariates explain about 75\% of within-site outcome variance, they explain only about 2\% here (Table~\ref{tab:dgp_params}), because the dummy slopes are set to moderate NJCS-scale contrasts. Thus, within-site covariate adjustment adds little. The Mixed-$\boldsymbol{X}$ DGP shows that weak covariates, not categorical ones, drive this result. Its covariates include a categorical one but remain strongly predictive, and the models fitted to individual-level data again have the lowest RMSE.

Consequently, site-Sum-OLS nearly coincides with Plug-in OLS, and Site-Sum-Bayes performs similarly to the individual-data BHMs. The sampling errors of the covariate means are also nearly uncorrelated with those of the outcome means, so modeling them jointly adds little. The main remaining difference among methods is how they handle estimation error in the control mean: the OLS comparators ignore it, 2S-ME shrinks it, and the Bayesian models integrate over it. The similar performance therefore reflects the weak covariates in this DGP rather than a limitation of modeling individual-level data.

\begin{figure}[!htbp]
    \centering
    \includegraphics[width=\linewidth]{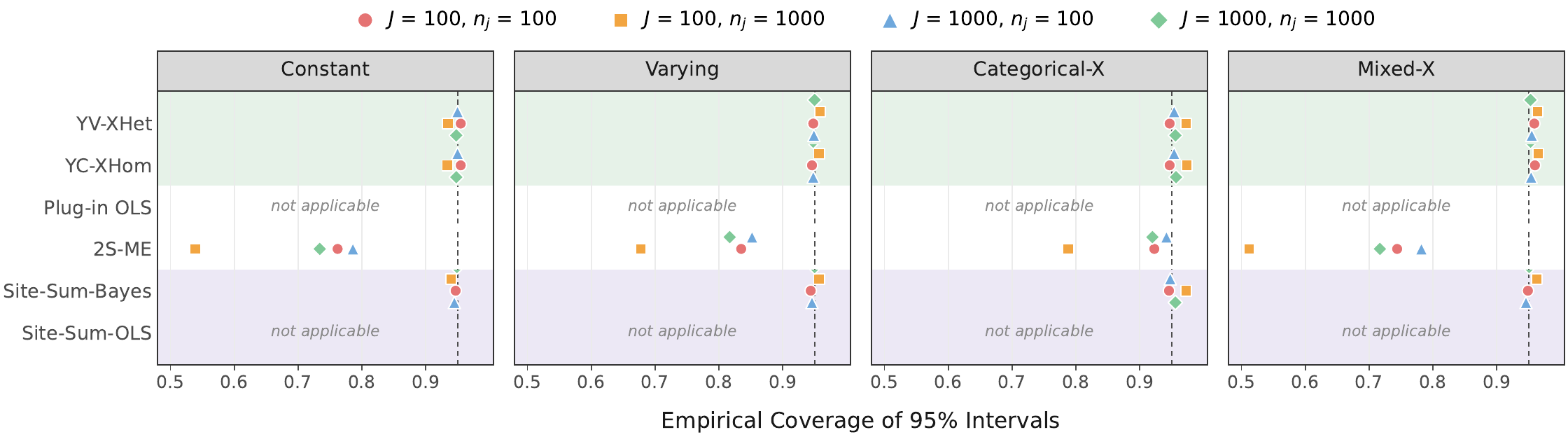}
    \caption{Empirical coverage of 95\% intervals for $\theta_j$. The dashed line indicates nominal coverage of 0.95. The two OLS comparators provide point estimates only. Markers are as in Figure~\ref{fig:rmse_rel}.}
    \label{fig:coverage}
\end{figure}

Figure~\ref{fig:coverage} compares the frequentist coverage of the interval estimates for the LRE. Plug-in OLS and Site-Sum-OLS provide no intervals because their LRE estimates are regression residuals without standard errors. The two individual-data BHMs and Site-Sum-Bayes have coverage close to the nominal 95\% level across DGPs and design cells. In contrast, 2S-ME under-covers under the Constant, Varying, and Mixed-$\boldsymbol{X}$ DGPs, with coverage of roughly 75--85\%, which falls to about 50--70\% when $J=100$ and $n_j=1{,}000$. Under the Categorical-$\boldsymbol{X}$ DGP, 2S-ME coverage is between 90\% and 95\% in all design cells except when $J=100$ and $n_j=1{,}000$, where it falls to slightly below 80\%.

This coverage difference between the Bayesian models and 2S-ME arises because the estimation uncertainty in the LRE has several sources. The estimated site-specific quantities carry sampling error that decreases with $n_j$, and the ecological regression coefficients $\boldsymbol{\beta}$ carry estimation uncertainty that decreases with $J$. The 2S-ME intervals account for sampling uncertainty in the site's random treatment effect estimate, but they treat the estimated site conditions and $\boldsymbol{\beta}$ as known. The Bayesian intervals account for all of these sources. Because the error in the estimated site-level conditions does not shrink as $J$ grows, 2S-ME coverage stays below the nominal level even with 1{,}000 sites. Under the Constant, Varying, and Mixed-$\boldsymbol{X}$ DGPs, coverage is poorest when sites are large but few ($J=100$, $n_j=1{,}000$). There, the intervals narrow as $n_j$ grows, whereas the ignored uncertainty in $\boldsymbol{\beta}$ remains large because $J$ is small. Under the Categorical-$\boldsymbol{X}$ DGP at $n_j=100$, the sampling error of the site's treatment effect estimate is large, so the two ignored sources matter less and 2S-ME coverage is close to nominal.

\begin{figure}[!htbp]
    \centering
    \includegraphics[width=\linewidth]{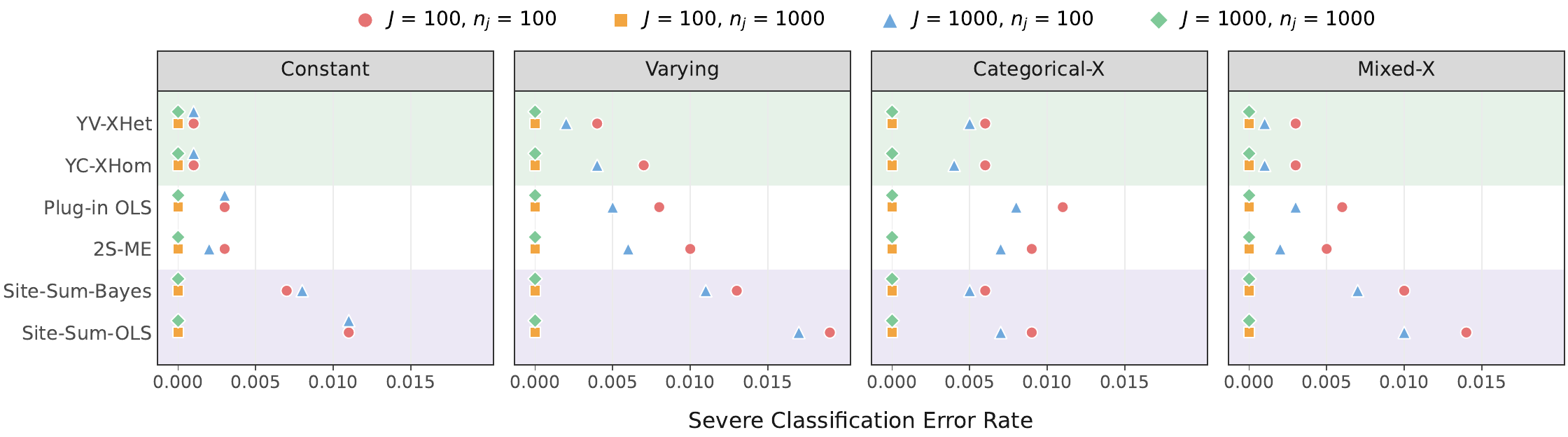}
    \caption{Severe classification error (SCE) rate: the proportion of sites assigned to the tail opposite their true tail. Marks are as in Figure~\ref{fig:rmse_rel}.}
    \label{fig:sce}
\end{figure}

\begin{figure}[!htbp]
    \centering
    \includegraphics[width=\linewidth]{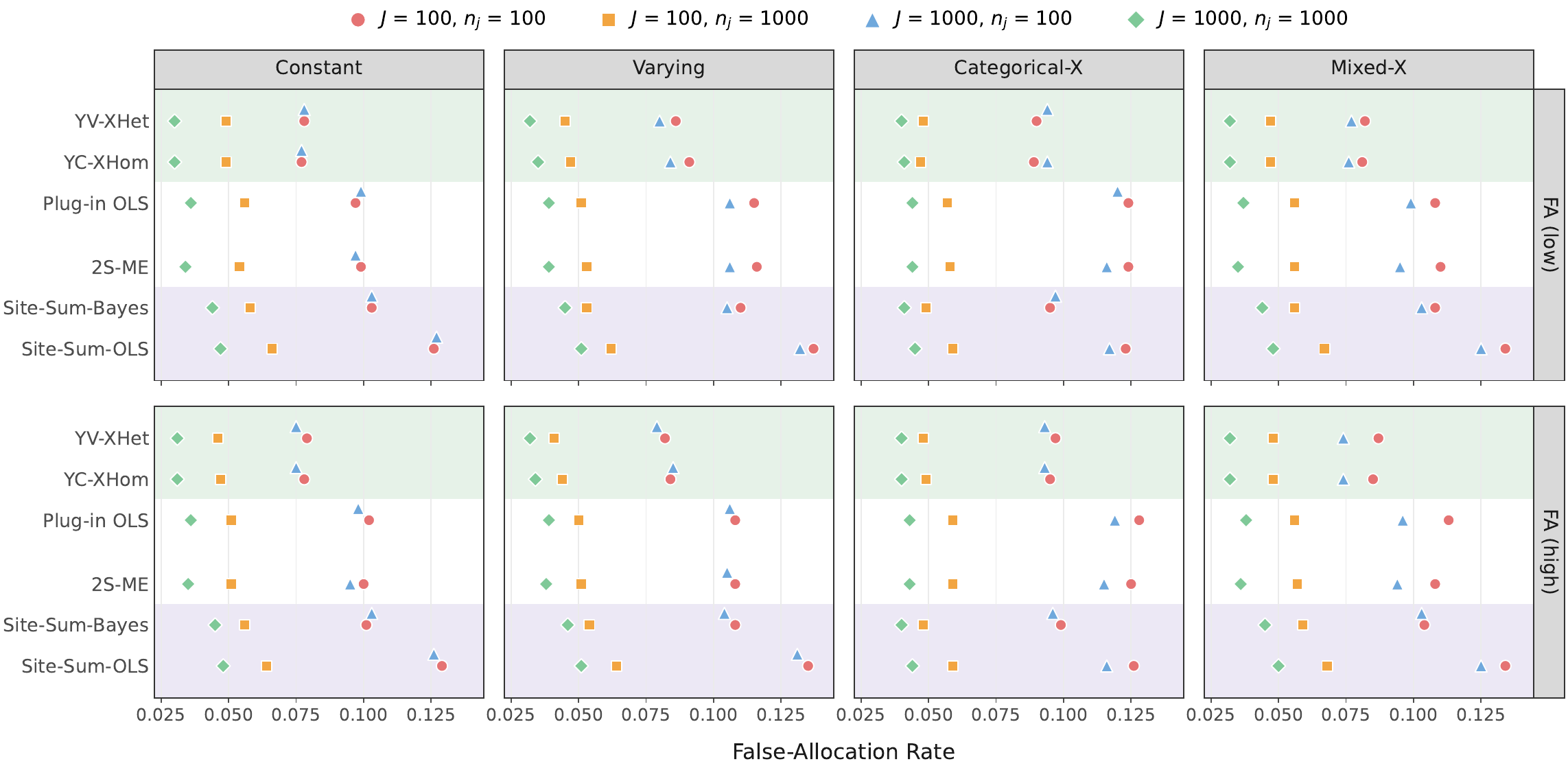}
    \caption{Rates of false-allocation into the lower tier ($\mathrm{FA}^{\mathrm{lower}}$, top panel) and upper tier ($\mathrm{FA}^{\mathrm{upper}}$, bottom panel). Markers are as in Figure~\ref{fig:rmse_rel}.}
    \label{fig:fa}
\end{figure}

Figures~\ref{fig:sce} and~\ref{fig:fa} show tier classification performance. Misclassification occurs when an organization is placed in the wrong tier due to LRE estimation error, and the classification error is severe when the organization is assigned to the opposite tail of the LRE distribution. As Figure~\ref{fig:sce} shows, severe classification errors are rare for all methods (below 2\% in every design cell) and fall to nearly zero at $n_j=1{,}000$. At $n_j=100$, the rankings of SCE rates under the Constant, Varying, and Mixed-$\boldsymbol{X}$ DGPs follow the RMSE pattern: the individual-data BHMs perform best, followed by 2S-ME and Plug-in OLS and then the site-level methods, with Site-Sum-OLS performing worst. Under the Categorical-$\boldsymbol{X}$ DGP, Site-Sum-Bayes performs like the individual-data BHMs and has lower SCE rates than the 2S-ME and OLS comparators.

False-allocation rates capture a more common error than SCE rates: assigning middle-tier organizations to a tail. Figure~\ref{fig:fa} shows clearer differences across methods. Under the Constant, Varying, and Mixed-$\boldsymbol{X}$ DGPs at $n_j=100$, the rates are about 8\% for the individual-data BHMs, about 10\% for Plug-in OLS, 2S-ME, and Site-Sum-Bayes, and about 13\% for Site-Sum-OLS. When evaluating the LREs of 100 organizations, the individual-data BHMs thus misallocate roughly one to two fewer centers to each tail than 2S-ME. At $n_j=1{,}000$, false-allocation rates decline and the differences between methods narrow. The false allocation rates are similar for the two tails. Under the Categorical-$\boldsymbol{X}$ DGP at $n_j=100$, Site-Sum-Bayes performs comparably to the individual-data BHMs, with rates of about 9\%, whereas 2S-ME, Plug-in OLS, and Site-Sum-OLS have rates near 12\%.

In summary, the results largely support the hypotheses in Section~\ref{sec:comparison}. The methods follow the expected RMSE ordering, except under Categorical-$\boldsymbol{X}$, where accuracy differences are small for the reasons discussed above. As expected, 2S-ME intervals under-cover, while Bayesian intervals have coverage close to nominal. The Bayesian models are also robust to misspecification. Under the Varying DGP, YC-XHom incorrectly assumes constant slopes, residual variances, and covariate covariance matrices across sites. This leads to a modest accuracy loss, largest at $n_j=100$, but has little effect on coverage or tier recovery. 
Under Categorical-$\boldsymbol{X}$ and Mixed-$\boldsymbol{X}$, the normality assumptions for within-site covariates and their site means are violated. The individual-data BHMs still keep coverage close to nominal, and their accuracy is at least comparable to the best alternative. When only site-level summaries are available, Site-Sum-Bayes provides intervals with coverage close to nominal, although with some loss of accuracy except under Categorical-$\boldsymbol{X}$. Overall, the accuracy gains of the individual-data BHMs are greatest for small sites, while their coverage advantages persist across sample sizes. Accounting for uncertainty in site-level conditions is therefore valuable even in trials with many sites.

Appendix~\ref{app:extra_results} presents complete simulation results for ten methods, which additionally include two individual-data BHMs that relax the constraints on either site-varying slopes and residual variances in the outcome model or on the within-site covariate covariance structure, and two additional 2S-ME variants with adjustments for shrinkage. For example, the results suggest that, under the varying DGP, the accuracy loss in YC-XHom comes from the constant slopes and residual variances rather than the shared covariance matrix. 

\section{Empirical Application}
\label{sec:application}

\subsection{Data Information}
The analytic sample includes 10,001 NJCS participants from 100 Job Corps centers (6,072 treatment and 3,929 control), excluding those without a center link or with missing outcomes. Center sizes range from 27 to 489 (median 85.5), so most centers are closer to the $n_j=100$ simulation setting. The outcome $Y_{ij}$ is average weekly earnings (in dollars) in the fourth year after random assignment. The six individual-level baseline covariates are age, an indicator for female, log-transformed prior-year earnings, months employed in the prior year, and indicators for Black and Hispanic participants. Two center-level covariates are the proportions of participants residing in superdense and not-dense ZIP codes, based on the NJCS classification of residential areas. These proportions are constant within centers and, together with the six individual-level covariates, form $\boldsymbol{X}_{ij}$, modeled as in the simulations. The site-level conditions are the control-group mean outcome $\mu_{0j}$ and the eight covariate means $\boldsymbol{\mu}_{Xj}$, estimated jointly with the LRE. Table~\ref{tab:njcs_desc} in Appendix~\ref{app:njcs_desc} summarizes center characteristics, which vary widely across centers. 

\subsection{Results of Estimation and Inference}

We report results from YV-XHet, the most flexible model, fitted with all eight covariates and allowing the slopes, residual variances, and covariate covariance matrices to vary across centers. The results are consistent across the other Bayesian models, namely YC-XHom, Site-Sum-Bayes, and two additional individual-data BHMs described in Appendix~\ref{app:other_models}. Their posterior mean LREs correlate at least 0.97 with those from YV-XHet, and at least 92 of 100 centers receive the same tier assignment. All models converged, with all $\hat{R}$ values below 1.05 and no divergent transitions.
\begin{figure}[!htbp]
\noindent\textit{(a) LRE}\\
\makebox[\linewidth][c]{\includegraphics[width=0.9\linewidth]{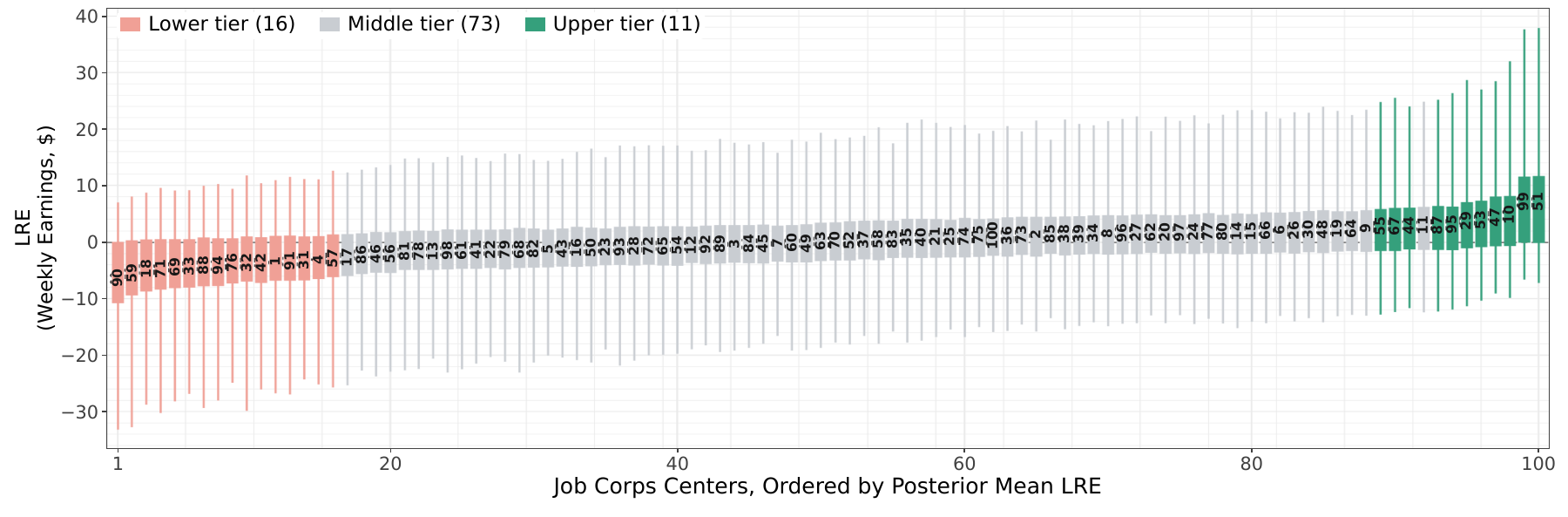}}\\[0.5em]
\textit{(b) ITT effect}\\
\makebox[\linewidth][c]{\includegraphics[width=0.9\linewidth]{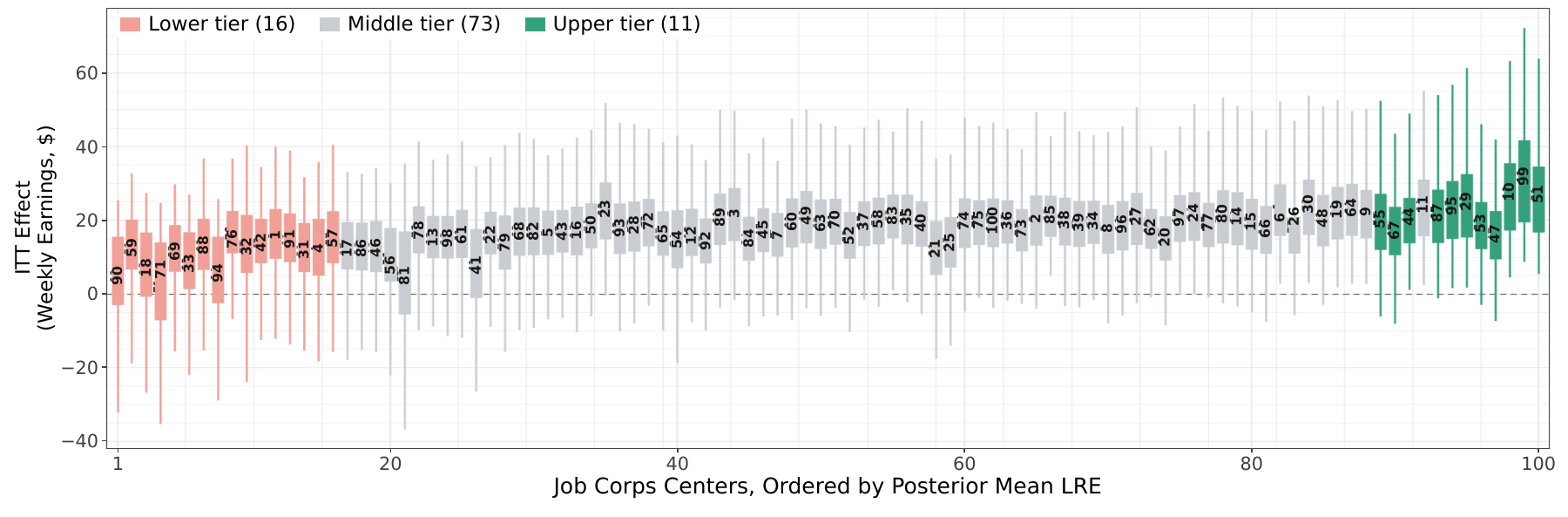}}
\caption{LRE (panel a) and ITT effect (panel b) for the 100 Job Corps centers, estimated with YV-XHet. Thick and thin bars are 50\% and 95\% credible intervals. Centers are labeled by number and ordered by posterior mean LRE in both panels.}
\label{fig:njcs_caterpillar}
\end{figure}
Figure~\ref{fig:njcs_caterpillar} compares the LREs and site-specific ITT effects for the 100 centers, ordered by posterior mean LRE in both panels. Panel (a) shows that most LRE estimates are close to zero, suggesting that centers generally perform near what would be expected under similar local conditions. Moreover, the wide and overlapping credible intervals make it difficult to distinguish most centers. Panel (b) shows that all posterior mean ITT effects are positive, with most around \$20 per week, although some centers have much larger effects than others. Compared with the ITT effects, the LREs vary less across centers. The posterior mean LREs have a standard deviation of \$2.5 across centers, about half that of the posterior mean ITT effects (5.0). This suggests that some of the differences in ITT effects may reflect differences in local conditions.

The ITT effects and LREs can lead to different conclusions about center performance. For example, Center 23 has one of the largest ITT effects but a slightly negative LRE, whereas Center 47 has a relatively small ITT effect but one of the highest LREs. Thus, rankings based on ITT effects alone may differ from those based on LRE. Nevertheless, the wide credible intervals indicate considerable uncertainty in both ITT and LRE rankings. Importantly, each LRE compares a center with centers under similar local conditions. A ranking of all 100 centers therefore mixes different benchmarks and should be read with the caveats of Section~\ref{sec:interpretation}.

\begin{figure}[!htbp]
\centering
\includegraphics[width=0.9\linewidth]{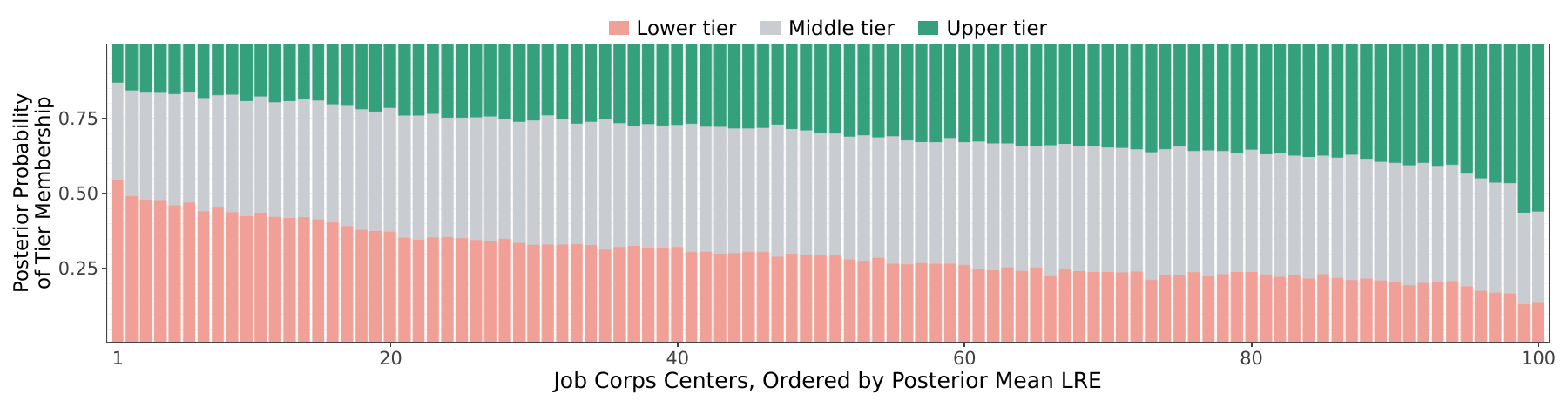}
\caption{Posterior probabilities of lower-, middle-, and upper-tier membership for the 100 Job Corps centers under YV-XHet. Each bar is a center, with three segments summing to one, ordered by posterior mean LRE.}
\label{fig:njcs_tier}
\end{figure}

To convey the uncertainty in these rankings, we compute each center's posterior probabilities of belonging to the lower, middle, or upper tier, as defined in Section~\ref{sec:setup_eval}. Figure~\ref{fig:njcs_tier} shows these probabilities for all 100 centers, ordered by posterior mean LRE. The probability of lower-tier membership generally decreases from left to right, while that of upper-tier membership increases. However, for most centers, there is considerable uncertainty about which tier they belong to. Only two centers have a probability greater than 0.5 of belonging to the upper tier, and only one has a probability greater than 0.5 of belonging to the lower tier. Assigning each center to its most probable tier yields 16 lower-tier, 73 middle-tier, and 11 upper-tier centers, as distinguished by the colors in Figure~\ref{fig:njcs_caterpillar}.

Overall, ITT effects and LREs can lead to different conclusions about center performance, and the posterior tier probabilities show how uncertain these comparisons are. Centers with relatively high probabilities of belonging to the upper tier may warrant further investigation to identify effective practices, while those more likely to belong to the lower tier may benefit from additional support.

\section{Discussion}
\label{sec:discussion}
Site-specific ITT effects reflect both organizational effectiveness and local conditions. LREs seek to distinguish the two by comparing each site's effect with the effect expected under its local conditions. The Bayesian hierarchical models developed here estimate site-specific LREs from individual-level data or site-level summaries, accounting for uncertainty in the site-level conditions. In simulations, the individual-data BHMs achieved the lowest estimation errors in nearly all settings, and all Bayesian models kept interval coverage close to nominal. The NJCS application illustrates how posterior tier probabilities reflect uncertainty in site rankings. As discussed in Section~\ref{sec:interpretation}, LRE comparisons across different local conditions require caution.

Several limitations warrant further investigation. The LRE estimated here adjusts for the local conditions captured by the control-group outcome mean and the covariate means. It equals the oracle LRE $\theta_j^{\ast}$ of Section~\ref{sec:estimand} only under Condition~\eqref{eq:lre_equiv}, that is, when unobserved conditions are related to the expected ITT effect only through these means. The two differ when, for example, an unobserved condition changes the ITT effect in a way that the control-group outcome mean does not reflect. Future research could examine the sensitivity of LRE estimates to such conditions.

The models also assume a linear relationship between expected ITT effects and local conditions and normally distributed site-level quantities. These assumptions performed well in our simulations, but some applications may need more flexibility. Local labor market conditions may relate nonlinearly to program effectiveness, and the proportion of Hispanic participants is highly skewed across NJCS centers. Nonparametric extensions could relax both assumptions. They would also allow the distribution of LRE to vary across local conditions, so that the standardized and quantile-transformed versions in Section~\ref{sec:interpretation} can be estimated and compared with the original LRE.

Another limitation concerns discrete and constant covariates. The individual-data BHMs treat binary and categorical covariates as normally distributed. Our simulations suggest that this has little effect on estimation accuracy, but modeling discrete covariates explicitly may represent the data better. In addition, some covariates are constant within a site, such as the female indicator at NJCS centers with no female participants, which makes the within-site covariance matrix singular. The individual-data BHMs add a small constant (0.01) to the diagonal of each within-site covariate covariance matrix to keep the matrices invertible. Site-Sum-Bayes instead treats the corresponding site-level means as known because their estimated sampling variances are zero. A unified treatment of such covariates would be a useful extension. Finally, the most flexible model YV-XHet required approximately four hours for the NJCS analysis, so faster estimation would help with larger trials.

Overall, the proposed Bayesian models enable  comparisons of organizational effectiveness that account for estimation uncertainty in site-level conditions and quantify uncertainty in relative rankings, yielding more accurate estimates and more valid inference than the existing methods we compared.

\section*{Acknowledgment}

This research was supported by a Spencer Foundation Major Research Grant (202400131). The views expressed are those of the authors, not the funding agency. We thank Howard Bloom, George Karabatsos, Stephen Raudenbush, Michael Sobel, and Doug Willms for insightful discussions, and the University of Chicago Research Computing Center (Midway3) and the University of Wisconsin--Madison Center for High Throughput Computing (CHTC) for computational resources and technical support \citep{CHTC_2006}.

\renewcommand{\bibsection}{\section*{References}}

\bibliographystyle{apacite}

\bibliography{references}

\appendix
\section{Proofs and Derivations}
\label{app:proofs}

\begin{proof}[Proof of Proposition~\ref{prop:ident}]
A parameter is identified if it can be written as a function of the distribution of the observed data. Let $P_j$ denote the distribution of $(Y_{ij}, Z_{ij}, \boldsymbol{X}_{ij})$ at site $j$. By ignorability and positivity, $\mu_{zj} = \mathbb{E}[Y_{ij}(z) \mid j] = \mathbb{E}[Y_{ij} \mid Z_{ij} = z, j]$ for $z \in \{0,1\}$, so $\mu_{0j}$ and $\delta_j = \mu_{1j} - \mu_{0j}$ are functions of $P_j$. So is $\boldsymbol{\mu}_{Xj} = \mathbb{E}[\boldsymbol{X}_{ij} \mid j]$. Hence $\boldsymbol{\phi}_j = (\delta_j, \mu_{0j}, \boldsymbol{\mu}_{Xj}^T)^T$ is a known function of $P_j$. Because sites are independent draws from a population of sites, the distribution of $P_j$ across sites is part of the distribution of the observed data, and it determines the joint distribution of $\boldsymbol{\phi}_j$. The conditional mean $\mathbb{E}[\delta_j \mid \mu_{0j}, \boldsymbol{\mu}_{Xj}]$ is a function of this joint distribution and exists because $\mathbb{E}|\delta_j| < \infty$. Therefore $\theta_j = \delta_j - \mathbb{E}[\delta_j \mid \mu_{0j}, \boldsymbol{\mu}_{Xj}]$ in Equation~\eqref{eq:lre_def} is identified. This argument requires neither a parametric assumption on the between-site distribution nor a limit in $n_j$ or $J$, as sample sizes affect estimation rather than identification.
\end{proof}

\begin{proof}[Proof of Lemma~\ref{lem:gauss}]
Partition the site-level vector $\boldsymbol{\phi}_j = (\delta_j, \boldsymbol{\phi}_{2j}^T)^T$ with mean $(\delta, \boldsymbol{m}_2^T)^T$ and the partition of $\boldsymbol{\Omega}$ in~\eqref{eq:phi_normal}. The Gaussian conditional of $\delta_j$ given $\boldsymbol{\phi}_{2j}$ has mean $\delta + \boldsymbol{\omega}_{21}^T \boldsymbol{\Omega}_{22}^{-1} (\boldsymbol{\phi}_{2j} - \boldsymbol{m}_2)$ and variance $\Omega_{11} - \boldsymbol{\omega}_{21}^T \boldsymbol{\Omega}_{22}^{-1}\boldsymbol{\omega}_{21}$, the Schur complement of $\boldsymbol{\Omega}_{22}$ in $\boldsymbol{\Omega}$. Setting the ecological slope $\boldsymbol{\beta} = \boldsymbol{\Omega}_{22}^{-1}\boldsymbol{\omega}_{21}$ gives the linear form of the conditional mean in the slope equation~\eqref{eq:lem_slope} and identifies the conditional variance with the cross-site variance $\sigma_\theta^2$ in the dispersion equation~\eqref{eq:sigma_theta}. Since $\theta_j$ is the conditional residual, it has mean zero, variance $\sigma_\theta^2$, and zero covariance with $\boldsymbol{\phi}_{2j}$.
\end{proof}

\begin{proof}[Derivation of the plug-in limit~\eqref{eq:plim_b} and the marginal~\eqref{eq:marginal_phihat}]
The across-site OLS slope is $\boldsymbol{b} = \widehat{\mathrm{Var}}(\hat{\boldsymbol{\phi}}_{2j})^{-1} \widehat{\mathrm{Cov}}(\hat{\boldsymbol{\phi}}_{2j}, \hat{\delta}_j)$. Because the sampling error has mean zero given $\boldsymbol{\phi}_j$ and is independent across sites, the law of large numbers gives $\widehat{\mathrm{Var}}(\hat{\boldsymbol{\phi}}_{2j}) \to_p \boldsymbol{\Omega}_{22} + \bar{\boldsymbol{\Sigma}}_{\mu\mu}$ and $\widehat{\mathrm{Cov}}(\hat{\boldsymbol{\phi}}_{2j}, \hat{\delta}_j) \to_p \boldsymbol{\omega}_{21} + \bar{\boldsymbol{\sigma}}_{\delta\mu}$, where the cross terms $\mathrm{Cov}(\boldsymbol{\phi}_{2j}, e_{\delta j})$ and $\mathrm{Cov}(\boldsymbol{e}_{\mu j}, \delta_j)$ vanish because the errors have conditional mean zero. This yields the plug-in limit~\eqref{eq:plim_b}. Each block of $\boldsymbol{\Sigma}_{\phi j}$ is a sampling covariance of within-site means and so is of order $n_j^{-1}$, giving the stated rate. The control mean is shared, with $\hat{\delta}_j = \bar{Y}_{1j} - \bar{Y}_{0j}$ and the two arm means computed on disjoint subsamples, so $\mathrm{Cov}(\hat{\delta}_j, \bar{Y}_{0j}) = -\mathrm{Var}(\bar{Y}_{0j}) = -v_{0j}/n_{0j}$, where $v_{0j}$ is the control-arm outcome variance (Appendix~\ref{app:site_cov}), the value produced by $\boldsymbol{\Xi} \hat{\boldsymbol{G}}_j \boldsymbol{\Xi}^T$ in the summary covariance~\eqref{eq:Sigma_j}.

For the marginal, integrating $\boldsymbol{\phi}_j$ in the measurement model~\eqref{eq:phihat_approx}, with the population $\boldsymbol{\Sigma}_{\phi j}$ in place of its estimate $\hat{\boldsymbol{\Sigma}}_{\phi j}$, over the between-site model~\eqref{eq:phi_normal} gives $\hat{\boldsymbol{\phi}}_j \sim \mathcal{N}(\boldsymbol{m}, \boldsymbol{\Omega} + \boldsymbol{\Sigma}_{\phi j})$ with $\boldsymbol{\Sigma}_{\phi j}$ known. The Gaussian likelihood in $\boldsymbol{m}$ and $\boldsymbol{\Omega}$ identifies $\boldsymbol{\Omega}$ by subtracting the known $\boldsymbol{\Sigma}_{\phi j}$ from the marginal covariance, so $\boldsymbol{\beta} = \boldsymbol{\Omega}_{22}^{-1}\boldsymbol{\omega}_{21}$ and $\sigma_\theta$ are recovered without the terms in~\eqref{eq:plim_b}.
\end{proof}

\section{Prior Specifications}
\label{app:priors}

Table~\ref{tab:prior_scales} reports the numerical values of every hyper-parameter in the Bayesian models. All five models share the between-site hyper-priors in \eqref{eq:prior_m}--\eqref{eq:prior_Omega} and differ in their within-site priors, as described in the Methods section. This appendix gives the rationale for the prior families and the prior centers.

\subsection*{Choice of Prior Families}

\textit{Half-normal on standard deviations.} For every standard-deviation parameter ($\tau_k$, $\tau_\alpha$, $\tau_{\log\sigma}$, $\sigma_z$, $\sigma_{Xk}$, $\tau_{\log\sigma_X}$), we use a half-normal hyper-prior $\mathcal{N}_+(0, s^2)$, following \citet{Gelman_2006}. The half-normal places adequate mass near zero, allowing a standard deviation to be small when the data suggest so. Half-Cauchy is a common alternative when the data carry little information about $\tau$, since its heavy tails admit large values (Polson \& Scott, 2012). In our design, $\tau$ is well identified by $J \geq 100$ sites. We therefore prefer the lighter tails of the half-normal, which yield more stable Hamiltonian Monte Carlo sampling (Stan Development Team, 2024).

\textit{LKJ$(2)$ on correlations.} All correlation matrices ($\boldsymbol{R}$, $\boldsymbol{R}_X$, $\boldsymbol{R}_{Xj}$) use the LKJ prior with shape $\eta = 2$. The LKJ$(\eta)$ family places a single parameter on correlation matrices. Setting $\eta = 1$ is uniform over correlation matrices, while $\eta > 1$ concentrates mass toward the identity \citep{Lewandowski_2009}. We set $\eta = 2$, which gently regularizes toward zero correlations while still allowing moderate correlations to emerge when the data support them (Stan Development Team, 2024).

\subsection*{Anchoring the Prior Means}

The prior mean $\boldsymbol{m}_0 = (13.2, 197.3, 0, 0)^T$ uses the between-site means of the ITT effect and control outcome from the simulation calibration. The covariate means are set to zero by the centering convention. The log-variance prior centers $m_{\log\sigma,0} = \log(8075)$ and $m_{\log\sigma,1} = \log(6459)$ use the residual-variance decomposition from the same calibration, detailed in Appendix~\ref{app:dgp}. The prior scales $\boldsymbol{s}_m$ and $s_{\mu_{\log\sigma}}$ are wide enough that the data dominate the posterior, as verified by a prior predictive check.

\begin{table}[H]
\caption{Hyper-parameter Values for the Bayesian Hierarchical Models}
\label{tab:prior_scales}
\begin{threeparttable}
\begin{tabularx}{\linewidth}{llX}
Symbol & Value & Role \\
\midrule
\multicolumn{3}{l}{\textit{Between-site hyper-priors (shared across all models)}} \\
$\boldsymbol{m}_0$ & $(13.2,\; 197.3,\; 0,\; 0)^T$ & Prior center of $\boldsymbol{m}$, NJCS-scale anchor \\
$\boldsymbol{s}_m$ & $(100,\; 300,\; 5,\; 5)^T$ & Prior SD of components of $\boldsymbol{m}$ \\
$\{s_k\}_{k=1}^{4}$ & $(100,\; 100,\; 5,\; 5)$ & Half-normal scale of $\tau_k$ \\
\midrule
\multicolumn{3}{l}{\textit{Within-site slopes and residual variances}} \\
$s_\alpha$ & 200 & Prior SD of $\bar{\boldsymbol{\alpha}}_z$ (YV) or $\boldsymbol{\alpha}_z$ (YC) \\
$s_{\tau_\alpha}$ & 50 & Half-normal scale of $\tau_\alpha$ (YV models) \\
$m_{\log\sigma,z}$ & $\log 8075,\; \log 6459$ & Prior center of $\mu_{\log\sigma,z}$, $z \in \{0,1\}$ (YV models) \\
$s_{\mu_{\log\sigma}}$ & 2 & Prior SD of $\mu_{\log\sigma,z}$ around $m_{\log\sigma,z}$ (YV models) \\
$s_{\tau_{\log\sigma}}$ & 1 & Half-normal scale of $\tau_{\log\sigma}$ (YV models) \\
$s_\sigma$ & 200 & Half-normal scale of $\sigma_z$ (YC models) \\
\midrule
\multicolumn{3}{l}{\textit{Within-site covariate covariance}} \\
$s_{\sigma_X}$ & 1 & Half-normal scale of $\sigma_{Xk}$ (XHom models) \\
$s_{\mu_X}$ & 1 & Prior SD of components of $\boldsymbol{\mu}_{\log\sigma_X}$ (XHet models) \\
$s_{\tau_X}$ & 0.5 & Half-normal scale of $\tau_{\log\sigma_X}$ (XHet models) \\
\bottomrule
\end{tabularx}
\begin{tablenotes}
\small
\item \textit{Note.} All correlation matrices use $\mathrm{LKJ}(2)$. The prior centers $m_{\log\sigma,z}$ equal the logs of the NJCS-scale residual variances in Appendix~\ref{app:dgp}. Site-Sum-Bayes uses the between-site hyper-priors; the within-site rows do not apply.
\end{tablenotes}
\end{threeparttable}
\end{table}

\section{Additional Individual-Data BHMs}
\label{app:other_models}

Two further individual-data BHMs combine the building blocks of YV-XHet and YC-XHom in Section~\ref{sec:indiv_models}. Both keep the between-site hyper-priors~\eqref{eq:prior_m}--\eqref{eq:prior_Omega}.

\begin{itemize}
\item \textbf{YV-XHom} (individual-level data, varying outcome slopes and residual variances, homogeneous covariate covariance matrix) retains the hierarchical priors on the within-site slopes~\eqref{eq:alpha_prior} and residual variances~\eqref{eq:sigma_prior} of YV-XHet. The covariate covariance matrix is shared across sites through the prior~\eqref{eq:sigmax_shared} of YC-XHom.

\item \textbf{YC-XHet} (individual-level data, constant outcome slopes and residual variances, heterogeneous covariate covariance matrices) retains the site-specific $\boldsymbol{\Sigma}_{Xj}$ with the prior~\eqref{eq:sigmax_prior}--\eqref{eq:sigmax_hyper} of YV-XHet. The slopes and residual standard deviations are constant across sites through the priors~\eqref{eq:v1_priors} of YC-XHom.
\end{itemize}

\noindent
Simulation results for these two models are reported in Appendix~\ref{app:extra_results}.

\section{Site-Sum-Bayes Model Covariance}
\label{app:site_cov}

This appendix derives $\hat{\boldsymbol{\Sigma}}_{\phi j} = \boldsymbol{\Xi}\hat{\boldsymbol{G}}_j\boldsymbol{\Xi}^T$ stated in the methods section. The two arm means $\bar{Y}_{1j}$ and $\bar{Y}_{0j}$ are computed on disjoint subsamples, so their covariance is zero. Their variances are $v_{zj}/n_{zj}$ for $z \in \{0, 1\}$, where $v_{zj} = \mathrm{Var}(Y_{ij} \mid Z_{ij} = z)$ is the within-arm outcome variance at site $j$. The full site mean $\bar{\boldsymbol{X}}_j$ pools individuals across arms,
\begin{equation}
  \bar{\boldsymbol{X}}_j = \frac{1}{n_j}\sum_{i:\, Z_{ij}=1}\boldsymbol{X}_{ij} + \frac{1}{n_j}\sum_{i:\, Z_{ij}=0}\boldsymbol{X}_{ij},
  \label{eq:Xbar_decomp}
\end{equation}
giving $\mathrm{Var}(\bar{\boldsymbol{X}}_j) = \boldsymbol{\Sigma}_{Xj}/n_j$. The decomposition also makes $\bar{\boldsymbol{X}}_j$ share the arm-$z$ subsample with $\bar{Y}_{zj}$,
\begin{equation}
  \mathrm{Cov}(\bar{Y}_{zj},\, \bar{\boldsymbol{X}}_j)
  = \frac{1}{n_{zj}\, n_j}\,\mathrm{Cov}\!\Bigl(\sum_{i:\, Z_{ij}=z}Y_{ij},\, \sum_{i:\, Z_{ij}=z}\boldsymbol{X}_{ij}\Bigr)
  = \frac{1}{n_j}\,\mathrm{Cov}(Y_{ij},\, \boldsymbol{X}_{ij} \mid Z_{ij}=z),
  \label{eq:cov_YzX}
\end{equation}
where contributions from the opposite arm vanish by independence, and the inner sum has $n_{zj}$ same-unit terms that cancel one factor of $n_{zj}$. The population block matrix is therefore
\begin{equation}
  \boldsymbol{G}_j =
  \begin{pmatrix}
    v_{1j}/n_{1j} & 0 & \boldsymbol{s}_{YX,1j}^T/n_j \\
    0 & v_{0j}/n_{0j} & \boldsymbol{s}_{YX,0j}^T/n_j \\
    \boldsymbol{s}_{YX,1j}/n_j & \boldsymbol{s}_{YX,0j}/n_j & \boldsymbol{\Sigma}_{Xj}/n_j
  \end{pmatrix},
  \quad
  \boldsymbol{s}_{YX,zj} = \mathrm{Cov}(Y_{ij},\, \boldsymbol{X}_{ij} \mid Z_{ij}=z),
  \label{eq:Gj_population}
\end{equation}
and replacing $v_{zj}$, $\boldsymbol{s}_{YX,zj}$, and $\boldsymbol{\Sigma}_{Xj}$ in \eqref{eq:Gj_population} with their sample estimators recovers $\hat{\boldsymbol{G}}_j$ in \eqref{eq:Gj_matrix}.

\section{Sufficient-Statistic Likelihood}
\label{app:suffstat}

The individual-data BHMs in the methods section are fitted to within-site summaries rather than to the full set of records. Subscript $zj$ denotes an arm within a site and subscript $j$ a whole site; in the code these are \texttt{y\_arm\_lpdf} and \texttt{x\_site\_lpdf}. This appendix shows that the within-site likelihood depends on the site data only through these summaries, and gives the expressions used in the Stan implementation. The summaries reproduce the individual-level likelihood without approximation. Writing the likelihood this way lets the sampler use a fixed set of per-site quantities in place of all $n_j$ individual records. Each evaluation then carries a cost that does not grow with the within-site sample size.

Fix a site $j$ and an arm $z \in \{0, 1\}$. The arm mean intercept $\mu_{zj}$ takes the value $\mu_{0j}$ for the control arm and $\mu_{1j} = \mu_{0j} + \delta_j$ for the treatment arm, both defined earlier. Under the outcome model in \eqref{eq:genmodel_Y}, each individual in arm $z$ satisfies $Y_{ij} \mid \boldsymbol{X}_{ij} \sim \mathcal{N}\bigl(\mu_{zj} + \boldsymbol{\alpha}_{zj}^T(\boldsymbol{X}_{ij} - \boldsymbol{\mu}_{Xj}),\, \sigma_{zj}^2\bigr)$. Each individual is observed under one arm only, so the observed-data likelihood factors across the two arms, and the off-diagonal element of $\boldsymbol{\Upsilon}_j$ does not enter. Only $\sigma_{0j}^2$ and $\sigma_{1j}^2$ appear below.

\subsection*{Arm-Level Outcome Likelihood}

Define the arm-level summaries from the $n_{zj}$ individuals in arm $z$. The arm means are $\bar{Y}_{zj} = n_{zj}^{-1}\sum_{i:\, Z_{ij}=z} Y_{ij}$ and $\bar{\boldsymbol{X}}_{zj} = n_{zj}^{-1}\sum_{i:\, Z_{ij}=z}\boldsymbol{X}_{ij}$. The arm-centered sums of squares and cross-products are
\begin{equation}
\begin{aligned}
  S^{YY}_{zj}
  &= \sum_{i:\, Z_{ij}=z}(Y_{ij} - \bar{Y}_{zj})^2, \\
  \boldsymbol{S}^{XY}_{zj}
  &= \sum_{i:\, Z_{ij}=z}(\boldsymbol{X}_{ij} - \bar{\boldsymbol{X}}_{zj})(Y_{ij} - \bar{Y}_{zj}), \\
  \boldsymbol{S}^{XX}_{zj}
  &= \sum_{i:\, Z_{ij}=z}(\boldsymbol{X}_{ij} - \bar{\boldsymbol{X}}_{zj})(\boldsymbol{X}_{ij} - \bar{\boldsymbol{X}}_{zj})^T,
\end{aligned}
\label{eq:arm_ss}
\end{equation}
a scalar, a $p$-vector, and a $p \times p$ matrix, each centered at the arm means and computed once from the data.

The arm log-likelihood sums $n_{zj}$ Gaussian log-densities,
\begin{equation}
\begin{aligned}
  \log L^{Y}_{zj}
  ={}& -\frac{n_{zj}}{2}\log(2\pi\sigma_{zj}^2) \\
   & - \frac{1}{2\sigma_{zj}^2}\sum_{i:\, Z_{ij}=z}\bigl(Y_{ij} - \mu_{zj} - \boldsymbol{\alpha}_{zj}^T(\boldsymbol{X}_{ij} - \boldsymbol{\mu}_{Xj})\bigr)^2.
\end{aligned}
\label{eq:arm_loglik_raw}
\end{equation}
To express the sum through the summaries in \eqref{eq:arm_ss}, add and subtract the arm means inside each residual. The residual then splits into a within-arm part and a constant part,
\begin{equation}
\begin{aligned}
  & Y_{ij} - \mu_{zj} - \boldsymbol{\alpha}_{zj}^T(\boldsymbol{X}_{ij} - \boldsymbol{\mu}_{Xj}) \\
  &\quad = \bigl[(Y_{ij} - \bar{Y}_{zj}) - \boldsymbol{\alpha}_{zj}^T(\boldsymbol{X}_{ij} - \bar{\boldsymbol{X}}_{zj})\bigr] + r_{zj},
\end{aligned}
\label{eq:resid_split}
\end{equation}
where
\begin{equation}
  r_{zj} = (\bar{Y}_{zj} - \mu_{zj}) - \boldsymbol{\alpha}_{zj}^T(\bar{\boldsymbol{X}}_{zj} - \boldsymbol{\mu}_{Xj})
  \label{eq:gap}
\end{equation}
is the gap between the observed arm mean and the mean the model predicts, equal to the average residual across the arm. The bracketed part in \eqref{eq:resid_split} has mean zero across the arm, so its cross-product with the constant $r_{zj}$ sums to zero. Squaring \eqref{eq:resid_split} and summing therefore gives
\begin{equation}
\begin{aligned}
  & \sum_{i:\, Z_{ij}=z}\bigl(Y_{ij} - \mu_{zj} - \boldsymbol{\alpha}_{zj}^T(\boldsymbol{X}_{ij} - \boldsymbol{\mu}_{Xj})\bigr)^2 \\
  &\quad = n_{zj}\, r_{zj}^2 + \mathrm{RSS}_{zj},
\end{aligned}
\label{eq:ss_decomp}
\end{equation}
with the within-arm residual sum of squares
\begin{equation}
  \mathrm{RSS}_{zj}
  = S^{YY}_{zj}
    - 2\,\boldsymbol{\alpha}_{zj}^T\boldsymbol{S}^{XY}_{zj}
    + \boldsymbol{\alpha}_{zj}^T\boldsymbol{S}^{XX}_{zj}\boldsymbol{\alpha}_{zj}.
  \label{eq:rss}
\end{equation}
Substituting \eqref{eq:ss_decomp} into \eqref{eq:arm_loglik_raw} gives the arm log-likelihood in summary form,
\begin{equation}
\begin{aligned}
  \log L^{Y}_{zj}
  ={}& -\frac{n_{zj}}{2}\log(2\pi\sigma_{zj}^2) \\
   & - \frac{n_{zj}\, r_{zj}^2 + \mathrm{RSS}_{zj}}{2\sigma_{zj}^2}.
\end{aligned}
\label{eq:arm_loglik_ss}
\end{equation}
The parameters enter only through $r_{zj}$ and $\mathrm{RSS}_{zj}$, and both are read from the arm means and the summaries in \eqref{eq:arm_ss}. Once these are fixed, the likelihood does not change with the individual records.

\subsection*{Site-Level Covariate Likelihood}

The covariate model in \eqref{eq:genmodel_X} applies to all $n_j$ individuals at the site, pooled across arms. Let $\bar{\boldsymbol{X}}_{j} = n_j^{-1}\sum_{i=1}^{n_j}\boldsymbol{X}_{ij}$ be the pooled site mean, and let
\begin{equation}
  \boldsymbol{Q}_{j}
  = \sum_{i=1}^{n_j}(\boldsymbol{X}_{ij} - \bar{\boldsymbol{X}}_{j})(\boldsymbol{X}_{ij} - \bar{\boldsymbol{X}}_{j})^T
  \label{eq:site_ss}
\end{equation}
be the pooled sum-of-squares matrix, centered at $\bar{\boldsymbol{X}}_{j}$. This matrix centers at the pooled site mean and so differs from the arm-centered $\boldsymbol{S}^{XX}_{zj}$ in \eqref{eq:arm_ss}; it is not the sum of those two arm matrices.

The covariate log-likelihood is
\begin{equation}
\begin{aligned}
  \log L^{X}_{j}
  ={}& -\frac{n_j p}{2}\log(2\pi)
       - \frac{n_j}{2}\log\det\boldsymbol{\Sigma}_{Xj} \\
   & - \frac{1}{2}\sum_{i=1}^{n_j}(\boldsymbol{X}_{ij} - \boldsymbol{\mu}_{Xj})^T\boldsymbol{\Sigma}_{Xj}^{-1}(\boldsymbol{X}_{ij} - \boldsymbol{\mu}_{Xj}).
\end{aligned}
\label{eq:x_loglik_raw}
\end{equation}
Write $\boldsymbol{h}_j = \bar{\boldsymbol{X}}_{j} - \boldsymbol{\mu}_{Xj}$ for the gap between the site mean and the parameter. Centering each term at $\bar{\boldsymbol{X}}_{j}$, the outer-product sum becomes
\begin{equation}
\begin{aligned}
  & \sum_{i=1}^{n_j}(\boldsymbol{X}_{ij} - \boldsymbol{\mu}_{Xj})(\boldsymbol{X}_{ij} - \boldsymbol{\mu}_{Xj})^T \\
  &\quad = \boldsymbol{Q}_{j} + n_j\, \boldsymbol{h}_j \boldsymbol{h}_j^T
   =: \boldsymbol{A}_j,
\end{aligned}
\label{eq:Aj}
\end{equation}
because the cross-terms with $\bar{\boldsymbol{X}}_{j}$ vanish. The quadratic sum in \eqref{eq:x_loglik_raw} equals $\mathrm{tr}(\boldsymbol{\Sigma}_{Xj}^{-1}\boldsymbol{A}_j)$, so the covariate log-likelihood reads
\begin{equation}
\begin{aligned}
  \log L^{X}_{j}
  ={}& -\frac{n_j p}{2}\log(2\pi)
       - \frac{n_j}{2}\log\det\boldsymbol{\Sigma}_{Xj} \\
   & - \frac{1}{2}\,\mathrm{tr}(\boldsymbol{\Sigma}_{Xj}^{-1}\boldsymbol{A}_j).
\end{aligned}
\label{eq:x_loglik_ss}
\end{equation}
The data enter only through $\bar{\boldsymbol{X}}_{j}$ and $\boldsymbol{Q}_{j}$.

\subsection*{Site Likelihood and Model Variants}

Combining the two arms with the covariate term, the log-likelihood contributed by site $j$ is
\begin{equation}
  \log L_j = \log L^{X}_{j} + \log L^{Y}_{0j} + \log L^{Y}_{1j},
  \label{eq:site_loglik}
\end{equation}
evaluated through \eqref{eq:arm_loglik_ss} and \eqref{eq:x_loglik_ss}. The four individual-data BHMs share this site likelihood and differ in how they treat the within-site parameters. The varying models place hierarchical priors on $\boldsymbol{\alpha}_{zj}$ and $\sigma_{zj}^2$. Under the constant models these are held fixed across sites. For the covariate covariance, the heterogeneous models give each site its own $\boldsymbol{\Sigma}_{Xj}$. The homogeneous models impose a single $\boldsymbol{\Sigma}_{X}$, so the determinant and inverse in \eqref{eq:x_loglik_ss} are formed once and reused across sites. The summaries in \eqref{eq:arm_ss} and \eqref{eq:site_ss} are computed before sampling, so each evaluation of \eqref{eq:site_loglik} costs $O(p^2)$ per site, independent of $n_j$. Sampling uses Stan's No-U-Turn sampler with a target acceptance rate of 0.95 (\texttt{adapt\_delta}) and a maximum tree depth of 12 (\texttt{max\_treedepth}). For each Bayesian model and replication, we run four chains with 1{,}000 warmup iterations (2{,}000 for $n_j=1{,}000$) and 2{,}500 post-warmup draws per chain. We monitor $\hat{R}$, effective sample sizes, divergent transitions, and E-BFMI, using thresholds of 1.01, 400, zero, and 0.3, respectively (Betancourt, 2017). The OLS comparators use \texttt{lm()}.

\subsection*{Parameterization of the Site-Level Layers}

Sampling efficiency in hierarchical models depends on how each site-level layer is parameterized. In the centered form, a site-level parameter is sampled directly from its between-site distribution, which is centered at the between-site mean. In the non-centered form, the sampler draws a standardized deviate, and the parameter is computed as the between-site mean plus the between-site standard deviation times that deviate. The two forms define the same model. The non-centered form samples efficiently when within-site data carry little information relative to the between-site distribution, and the centered form when within-site information is strong (Betancourt \& Girolami, 2015; Papaspiliopoulos et al., 2007). A mismatch shows up as divergent transitions, low energy fraction (E-BFMI), and inflated $\hat{R}$ (Betancourt, 2017). Following this guidance, we match the parameterization to each layer and design cell. All site-level layers use the non-centered form when $n_j = 100$. When $n_j = 1{,}000$, layers with positive true between-site variance use the centered form. Layers whose true between-site variance is zero (the slope and residual-variance layers under the constant-structure DGPs) remain non-centered at both sample sizes. The data then favor a between-site standard deviation near zero, where the centered form samples poorly (Betancourt \& Girolami, 2015). A pilot run of one replication per cell under this scheme produced no divergent transitions, $\hat{R} \leq 1.003$, and E-BFMI above 0.60 for every model, DGP, and cell.

\section{DGP Parameter Specification}
\label{app:dgp}
Table~\ref{tab:dgp_params} reports all numerical values used in the simulation. The between-site mean vector $\boldsymbol{m}$ and covariance matrix $\boldsymbol{\Omega}$ are calibrated to NJCS-derived relationships among $(\delta_j, \mu_{0j}, \boldsymbol{\mu}_{Xj})$, with calibration formulas in the code repository. The within-site error structure in the Constant DGP yields $\sigma_0 = \sqrt{8075} \approx 89.86$ and $\sigma_1 = \sqrt{6459} \approx 80.37$. For the two covariate-violation DGPs, the categorical covariates are labeled age and baseline earnings. Their category shares are set to those of three-category NJCS covariates, and the labels are illustrative. For age, the three groups 16--17, 18--19, and 20--24 have proportions $(0.437,\, 0.307,\, 0.256)$, reproduced by cutoffs $a_{11} = \Phi^{-1}(0.437) = -0.159$ and $a_{12} = \Phi^{-1}(0.744) = 0.656$, where $\Phi^{-1}$ is the standard normal quantile function. For baseline earnings, the three groups low, medium, and high have proportions $(0.469,\, 0.355,\, 0.176)$, reproduced by cutoffs $a_{21} = \Phi^{-1}(0.469) = -0.078$ and $a_{22} = \Phi^{-1}(0.824) = 0.931$. For age the threshold construction is a device for generating multinomial draws; for baseline earnings the latent continuum has a natural interpretation as the underlying earnings level. Each categorical covariate enters through two dummy indicators contrasting the second and third categories against the first, and the dummy slopes are set to moderate NJCS-scale contrasts (Table~\ref{tab:dgp_params}). Under Categorical-$\boldsymbol{X}$, the latent variables are centered at the site-varying means $\mu_{Xkj}$, so category composition varies across sites. In the NJCS, client composition varies markedly across sites (for example, a site-level SD of 29.5 percentage points for the share of Black participants), and this design carries comparable variation into the dummies. Under Mixed-$\boldsymbol{X}$, the third latent coordinate is centered at a site-varying mean $\mu_{X3j}$, which loads on the same between-site factor structure as $\mu_{X1j}$ and $\mu_{X2j}$ (extending $\boldsymbol{m}$ and $\boldsymbol{\Omega}$ by one coordinate with $\mathrm{Var}(\mu_{X3j}) = 0.10$), and correlates at $\rho = 0.3$ with both continuous covariates within site. Composition therefore varies across sites and covariate types are dependent. Under Categorical-$\boldsymbol{X}$, the residual variances are four times the Constant values ($\sigma_0^2 = 32{,}300$, $\sigma_1^2 = 25{,}836$, with the cross-arm covariance scaled by the same factor), chosen so that the raw site ITT estimates have reliability of about 0.55 at $n_j = 100$; at the Constant noise level, the discrete covariates leave the site-summary methods with little estimation error to correct, and the comparison among them is uninformative.

\begin{table}[H]
\caption{DGP Parameter Values}
\label{tab:dgp_params}
\begin{threeparttable}
\small
\setlength{\tabcolsep}{4pt}
\begin{tabularx}{\linewidth}{>{\raggedright\arraybackslash}p{0.27\linewidth} >{\raggedright\arraybackslash}p{0.21\linewidth} X}

Symbol & Value & Description \\
\midrule
$\delta$ & 13.2 & Overall mean ITT effect $\mathbb{E}[\delta_j]$ \\
$\mu_0$ & 197.3 & Overall mean control outcome $\mathbb{E}[\mu_{0j}]$ \\
$\mu_{X1}$, $\mu_{X2}$ & 0, 0 & Overall covariate means \\
\midrule
$[\boldsymbol{\Omega}]_{11}$ & 2488.44 & $\mathrm{Var}(\delta_j)$ \\
$[\boldsymbol{\Omega}]_{22}$ & 1622.15 & $\mathrm{Var}(\mu_{0j})$ \\
$[\boldsymbol{\Omega}]_{33}$, $[\boldsymbol{\Omega}]_{44}$ & 0.10, 0.10 & $\mathrm{Var}(\mu_{X1j})$, $\mathrm{Var}(\mu_{X2j})$ \\
$[\boldsymbol{\Omega}]_{12}$ & $-1159.54$ & $\mathrm{Cov}(\delta_j, \mu_{0j})$ \\
$[\boldsymbol{\Omega}]_{13}$, $[\boldsymbol{\Omega}]_{14}$ & $-5.19$, $4.88$ & $\mathrm{Cov}(\delta_j, \mu_{X1j})$, $\mathrm{Cov}(\delta_j, \mu_{X2j})$ \\
$[\boldsymbol{\Omega}]_{23}$, $[\boldsymbol{\Omega}]_{24}$ & $2.41$, $-3.25$ & $\mathrm{Cov}(\mu_{0j}, \mu_{X1j})$, $\mathrm{Cov}(\mu_{0j}, \mu_{X2j})$ \\
$[\boldsymbol{\Omega}]_{34}$ & 0 & $\mathrm{Cov}(\mu_{X1j}, \mu_{X2j})$ \\
\midrule
$\bar{\boldsymbol{\alpha}}_0$ & $(120,\; -100)^T$ & Mean within-site slopes, control group \\
$\bar{\boldsymbol{\alpha}}_1$ & $(124,\; -97.5)^T$ & Mean within-site slopes, treatment group \\
$\sigma_0$ & 89.86 & Control-group error SD (Constant DGP) \\
$\sigma_1$ & 80.37 & Treatment-group error SD (Constant DGP) \\
$\sigma_\theta$ & 38.15 & True LRE between-site SD (38.12 under Mixed-$\boldsymbol{X}$) \\
\midrule
$\tau_\alpha$ & 50 & Between-site SD of slopes (Varying-Slope and Varying DGPs) \\
$\sigma_0^2$, $\sigma_1^2$ & 8075, 6459 & Arm-specific residual variances; $\mathbb{E}[\sigma_{zj}^2] = \sigma_z^2$ under the Varying-Slope and Varying DGPs \\
$\tau_{\log\sigma}$ & 0.1 & Between-site SD of log-variance (Varying-Slope and Varying DGPs) \\
\midrule
$(a_{11},\, a_{12})$ & $(-0.159,\; 0.656)$ & Cutoffs, age (shares $0.437 / 0.307 / 0.256$) \\
$(a_{21},\, a_{22})$ & $(-0.078,\; 0.931)$ & Cutoffs, baseline earnings (shares $0.469 / 0.355 / 0.176$) \\
$\boldsymbol{\gamma}_0$, Categorical-$\boldsymbol{X}$ & $(20,\; 45,\; 30,\; -20)^T$ & Dummy slopes, control group (age 18--19, age 20--24, earnings medium, earnings high) \\
$\boldsymbol{\gamma}_1$, Categorical-$\boldsymbol{X}$ & $(22,\; 48,\; 32,\; -18)^T$ & Dummy slopes, treatment group \\
$\boldsymbol{\gamma}_0$, Mixed-$\boldsymbol{X}$ & $(30,\; -20)^T$ & Dummy slopes, control group (earnings medium, high vs.\ low) \\
$\boldsymbol{\gamma}_1$, Mixed-$\boldsymbol{X}$ & $(32,\; -18)^T$ & Dummy slopes, treatment group \\
$[\boldsymbol{\Omega}]_{55}$, Mixed-$\boldsymbol{X}$ & 0.10 & $\mathrm{Var}(\mu_{X3j})$, latent mean of the categorical covariate \\
$[\boldsymbol{\Omega}]_{15}$, $[\boldsymbol{\Omega}]_{25}$, Mixed-$\boldsymbol{X}$ & $-5.15$, $9.10$ & $\mathrm{Cov}(\delta_j, \mu_{X3j})$, $\mathrm{Cov}(\mu_{0j}, \mu_{X3j})$; $[\boldsymbol{\Omega}]_{35} = [\boldsymbol{\Omega}]_{45} = 0$ \\
$\rho$ & 0.3 & Within-site latent correlation with each continuous covariate (Mixed-$\boldsymbol{X}$) \\
\bottomrule
\end{tabularx}
\begin{tablenotes}
\small
\item \textit{Note.} Numerical values are calibrated to NJCS weekly-earnings outcomes. $[\boldsymbol{\Omega}]_{kk'}$ denotes the $(k,k')$ element of $\boldsymbol{\Omega}$; the block partition of Section~\ref{sec:genmodel} is not used here. $\sigma_\theta$ is computed from $\boldsymbol{\Omega}$ via Equation~\eqref{eq:sigma_theta}. For the Varying DGP, within-site covariate variances are drawn from $\mathrm{LogNormal}(0, 0.5^2)$ and correlations from $\mathrm{Uniform}(-0.5, 0.5)$, independently across sites. Under Categorical-$\boldsymbol{X}$, both latent variables are centered at the site-varying means $\mu_{Xkj}$ with unit variance. Under Mixed-$\boldsymbol{X}$, the continuous pair follows the Constant DGP and the categorical covariate is generated from a third latent coordinate with site-varying mean $\mu_{X3j}$, unit variance, and correlation $\rho$ with each continuous coordinate.
\end{tablenotes}
\end{threeparttable}
\end{table}

\section{Generative Model by DGP}
\label{app:dgp_models}

This appendix writes out the complete generative model for each DGP. Two layers are shared by all five. The DGPs differ only in the covariate model and in which within-site parameters vary across sites. All numerical values appear in Table~\ref{tab:dgp_params}.

\subsection*{Shared Layers}

The between-site layer draws the site-level vector $\boldsymbol{\phi}_j$, comprising the site ITT effect $\delta_j$, the control mean $\mu_{0j}$, and the $p^{*}$-vector of latent covariate means $\boldsymbol{\mu}^{*}_{Xj}$ ($p^{*} = 2$ for the first four DGPs, and $p^{*} = 3$ under Mixed-$\boldsymbol{X}$, whose third coordinate $\mu_{X3j}$ is the site mean of the categorical covariate's latent variable),
\begin{equation}
\boldsymbol{\phi}_j = (\delta_j,\; \mu_{0j},\; \boldsymbol{\mu}^{*T}_{Xj})^T
\overset{\mathrm{iid}}{\sim} \mathcal{N}_{p^{*}+2}(\boldsymbol{m},\, \boldsymbol{\Omega}),
\qquad j = 1, \ldots, J,
\label{eq:appmod_phi}
\end{equation}
with $\boldsymbol{m}$ and $\boldsymbol{\Omega}$ extended by one row and column under Mixed-$\boldsymbol{X}$, calibrated in the same way as the continuous covariate means. Treatment is assigned independently with probability 0.5 within each site. Potential outcomes in every DGP take the form $Y_{ij}(z) = f_z(\boldsymbol{X}_{ij}) + \varepsilon_{zij}$ for $z \in \{0, 1\}$, where the outcome function $f_z$ is specified per DGP below.
The residual layer draws the pair of arm-specific errors ($\varepsilon_{0ij}$ under control, $\varepsilon_{1ij}$ under treatment) jointly for each individual,
\begin{equation}
(\varepsilon_{0ij},\, \varepsilon_{1ij})^T
\overset{\mathrm{iid}}{\sim} \mathcal{N}_2\!\left(\boldsymbol{0},\;
\boldsymbol{\Upsilon}_j\right),
\qquad
\boldsymbol{\Upsilon}_j = \begin{pmatrix} \sigma_{0j}^2 & \sigma_{01} \\ \sigma_{01} & \sigma_{1j}^2 \end{pmatrix},
\qquad \sigma_{01} = 4{,}160 \ (16{,}640 \text{ under Categorical-}\boldsymbol{X}),
\label{eq:appmod_eps}
\end{equation}
where $\boldsymbol{\Upsilon}_j$ is the residual covariance of the potential outcomes from Section~\ref{sec:genmodel}. Its off-diagonal $\sigma_{01}$ makes the two potential outcomes positively correlated within person. It does not enter the observed-data likelihood, since $Y_{ij} = (1 - Z_{ij})\, Y_{ij}(0) + Z_{ij}\, Y_{ij}(1)$ reveals one error per individual.

\subsection*{Constant DGP}

Covariates are bivariate normal around their site means, and all within-site parameters are fixed across sites:
\begin{align}
\boldsymbol{X}_{ij} &\overset{\mathrm{iid}}{\sim} \mathcal{N}_2(\boldsymbol{\mu}_{Xj},\, I_2), \\
Y_{ij}(z) &= \mu_{0j} + z\,\delta_j + (\boldsymbol{X}_{ij} - \boldsymbol{\mu}_{Xj})^T \bar{\boldsymbol{\alpha}}_z + \varepsilon_{zij},
\qquad z \in \{0, 1\},
\label{eq:appmod_const}
\end{align}
with slopes $\bar{\boldsymbol{\alpha}}_0 = (120,\, -100)^T$, $\bar{\boldsymbol{\alpha}}_1 = (124,\, -97.5)^T$ and variances $\sigma_{zj}^2 = \sigma_z^2$.

\subsection*{Varying-Slope DGP}

The outcome equation is \eqref{eq:appmod_const} with site-specific slopes $\boldsymbol{\alpha}_{zj}$ in place of $\bar{\boldsymbol{\alpha}}_z$. Slopes and log residual variances vary across sites,
\begin{align}
\alpha_{zjk} &\overset{\mathrm{ind}}{\sim} \mathcal{N}(\bar{\alpha}_{zk},\, \tau_\alpha^2), \qquad k = 1, 2,\\
\log \sigma_{zj}^2 &\overset{\mathrm{ind}}{\sim} \mathcal{N}\!\left(\log \sigma_z^2 - \tfrac{1}{2}\tau_{\log\sigma}^2,\; \tau_{\log\sigma}^2\right),
\end{align}
where $\tau_\alpha$ and $\tau_{\log\sigma}$ are the between-site standard deviations of the slopes and of the log variance. The offset $-\tfrac{1}{2}\tau_{\log\sigma}^2$ keeps $\mathbb{E}[\sigma_{zj}^2] = \sigma_z^2$. Covariates remain $\mathcal{N}_2(\boldsymbol{\mu}_{Xj},\, I_2)$.

\subsection*{Varying DGP}

All layers follow the Varying-Slope DGP, and the covariate covariance also varies across sites,
\begin{equation}
\boldsymbol{X}_{ij} \overset{\mathrm{iid}}{\sim} \mathcal{N}_2(\boldsymbol{\mu}_{Xj},\, \boldsymbol{\Sigma}_{Xj}),
\qquad
\boldsymbol{\Sigma}_{Xj} = \begin{pmatrix} \sigma_{Xj1}^2 & \rho_{Xj}\,\sigma_{Xj1}\sigma_{Xj2} \\ \rho_{Xj}\,\sigma_{Xj1}\sigma_{Xj2} & \sigma_{Xj2}^2 \end{pmatrix},
\end{equation}
with variances $\sigma_{Xjk}^2 \sim \mathrm{LogNormal}(0,\, 0.5^2)$ and correlation $\rho_{Xj} \sim \mathrm{Uniform}(-0.5,\, 0.5)$, independently across sites.

\subsection*{Categorical-$\boldsymbol{X}$ DGP}

Both covariates are three-category. For covariate $X_k$ indexed by $k$ ($k = 1$ age; $k = 2$ baseline earnings), a latent normal variable centered at the site mean is thresholded at fixed cutoffs $(a_{k1}, a_{k2})$,
\begin{align}
X^{*}_{ijk} &\overset{\mathrm{iid}}{\sim} \mathcal{N}(\mu_{Xkj},\, 1), \qquad
C_{ijk} = c \;\; \text{iff} \;\; X^{*}_{ijk} \in (a_{k,c-1},\, a_{k,c}], \quad c \in \{1, 2, 3\},
\end{align}
with $a_{k0} = -\infty$ and $a_{k3} = \infty$. Each covariate yields two dummy indicators $\boldsymbol{B}_{ijk} = \big(\mathbf{1}(C_{ijk} = 2),\, \mathbf{1}(C_{ijk} = 3)\big)^T$, and the observed covariate vector stacks them, $\boldsymbol{X}_{ij} = (\boldsymbol{B}_{ij1}^T,\, \boldsymbol{B}_{ij2}^T)^T \in \{0,1\}^4$. Writing $\boldsymbol{\nu}_{jk} = \mathbb{E}[\boldsymbol{B}_{ijk} \mid j]$ for the site-true dummy expectations (available in closed form from the normal distribution function), the potential outcomes are
\begin{equation}
Y_{ij}(z) = \mu_{0j} + z\,\delta_j
+ \sum_{k=1}^{2} (\boldsymbol{B}_{ijk} - \boldsymbol{\nu}_{jk})^T \boldsymbol{\gamma}_{zk}
+ \varepsilon_{zij},
\label{eq:appmod_cat}
\end{equation}
where $\boldsymbol{\gamma}_{zk}$ collects the fixed arm-specific dummy slopes for covariate $k$; Table~\ref{tab:dgp_params} reports the stacked vector $\boldsymbol{\gamma}_z = (\boldsymbol{\gamma}_{z1}^T,\, \boldsymbol{\gamma}_{z2}^T)^T$. Centering the dummies at $\boldsymbol{\nu}_{jk}$ keeps $\mathbb{E}[Y_{ij}(0) \mid j] = \mu_{0j}$, so the between-site layer retains its meaning. Slopes and residual variances are constant across sites, as in the Constant DGP. Because $\mu_{Xkj}$ varies across sites, the category shares $\boldsymbol{\nu}_{jk}$ vary across sites.

The LRE in~\eqref{eq:lre_def} conditions on the site means of the observed dummy indicators, $\boldsymbol{\nu}_{jk}$. For each covariate, these site means are a one-to-one function of the latent mean $\mu_{Xkj}$, so conditioning on them is equivalent to conditioning on the latent means. The true $\theta_j$ is therefore the same under either conditioning. The residual variances are four times those of the Constant DGP (Table~\ref{tab:dgp_params}). Dummy covariates explain far less of the outcome variance than continuous ones, so without this increase the outcomes would be less variable and the site means more precise than under the other DGPs. The increase keeps the sampling error of the site-level summaries comparable across DGPs. The covariate covariance also differs across sites under this DGP, although no site-specific covariance is drawn. The variance of a dummy equals $\nu(1-\nu)$, where $\nu$ is its site mean, so sites with different proportions have different variances.

\subsection*{Mixed-$\boldsymbol{X}$ DGP}

All covariate information originates from one latent Gaussian vector per individual,
\begin{equation}
\boldsymbol{X}^{*}_{ij} = (X^{*}_{ij1},\, X^{*}_{ij2},\, X^{*}_{ij3})^T
\overset{\mathrm{iid}}{\sim} \mathcal{N}_3\!\left(\boldsymbol{\mu}^{*}_{Xj},\;
\boldsymbol{\Sigma}_\rho\right),
\qquad
\boldsymbol{\mu}^{*}_{Xj} = (\mu_{X1j},\, \mu_{X2j},\, \mu_{X3j})^T,
\qquad
\boldsymbol{\Sigma}_\rho = \begin{pmatrix} 1 & 0 & \rho \\ 0 & 1 & \rho \\ \rho & \rho & 1 \end{pmatrix},
\label{eq:appmod_mix_lat}
\end{equation}
with $\rho = 0.3$. The first two coordinates are observed directly as the continuous covariates, $X_{ijk} = X^{*}_{ijk}$ for $k = 1, 2$. The third coordinate is observed only through its category: thresholding $X^{*}_{ij3}$ at the baseline-earnings cutoffs gives the category $C_{ij3} \in \{1, 2, 3\}$ and the dummy pair $\boldsymbol{B}_{ij} = \big(\mathbf{1}(C_{ij3} = 2),\, \mathbf{1}(C_{ij3} = 3)\big)^T$, with site-specific expectations $\boldsymbol{\nu}_j = \mathbb{E}[\boldsymbol{B}_{ij} \mid j]$ determined by $\mu_{X3j}$ and the cutoffs. The observed covariate vector is $\boldsymbol{X}_{ij} = (X_{ij1},\, X_{ij2},\, \boldsymbol{B}_{ij}^T)^T$, and the potential outcomes are
\begin{equation}
Y_{ij}(z) = \mu_{0j} + z\,\delta_j
+ (X_{ij1} - \mu_{X1j},\, X_{ij2} - \mu_{X2j})\, \bar{\boldsymbol{\alpha}}_z
+ (\boldsymbol{B}_{ij} - \boldsymbol{\nu}_j)^T \boldsymbol{\gamma}_z
+ \varepsilon_{zij}.
\label{eq:appmod_mix}
\end{equation}
Slopes and residual variances are constant across sites. Because $\mu_{X3j}$ varies across sites, the composition $\boldsymbol{\nu}_j$ varies across sites and carries between-site signal about $\delta_j$, while $\rho$ makes the continuous and categorical covariates dependent within site.

\subsection*{A General Form}

The five covariate models are instances of one construction. Let $\boldsymbol{X}^{*}_{ij} \sim \mathcal{N}_{p^{*}}(\boldsymbol{\mu}^{*}_{Xj},\, \boldsymbol{\Sigma}^{*}_j)$ be a latent Gaussian vector, and let each observed covariate be either a latent coordinate itself or its thresholded category. The first three DGPs observe all coordinates directly ($p^{*} = 2$, with $\boldsymbol{\Sigma}^{*}_j$ equal to $I_2$ or site-varying). Categorical-$\boldsymbol{X}$ observes both coordinates only through their categories ($p^{*} = 2$, $\boldsymbol{\Sigma}^{*}_j = I_2$). Mixed-$\boldsymbol{X}$ observes two coordinates directly and one through its category ($p^{*} = 3$, $\boldsymbol{\Sigma}^{*}_j = \boldsymbol{\Sigma}_\rho$). Which coordinates are thresholded, and whether their latent means vary across sites, determines which violation each DGP creates. 

\section{Comparison Methods: \textsf{R} Implementation}
\label{app:comparison_methods}

This appendix specifies the comparison methods in the simulation. Three are base procedures (Site-Sum-OLS, Plug-in OLS, and 2S-ME), and two apply posterior-summary adjustments to the 2S-ME estimates (2S-ME-CB and 2S-ME-TG). Each procedure is presented through its \textsf{R} implementation, with brief in-code annotations indicating what each step computes. Code examples use the two-covariate setting ($p = 2$) of the simulation; the same structure extends naturally to additional covariates ($p \neq 2$). The OLS variants are standard two-stage regressions, and the 2S-ME procedure follows \citet{Hong2026AJE}. The two adjustments follow Louis (1984), Ghosh (1992), and Shen and Louis (1998). The 2S-ME estimates are empirical Bayes predictions shrunk toward the grand mean and therefore understate the between-site variation, which the adjustments are designed to restore. \citet{Lee_2025} study both adjustments for site-specific effects in multi-site trials; we apply them to the residual LRE. To name the estimands by what they are rather than reuse $\delta$ and $\theta$ from the proposed model, the code labels each site-specific ITT estimate \texttt{itt} and each LRE estimate \texttt{lre}.

\subsection*{Shared Data Preparation}

All three base methods operate on the same underlying inputs. The individual-level data frame \texttt{dat} contains columns \texttt{Y} (observed outcome), \texttt{Z} (treatment indicator), \texttt{j} (site index), and \texttt{X1}, \texttt{X2} (individual-level pre-treatment covariates). From this we derive the site-level summary \texttt{site\_sum} with one row per site.
\begin{tcolorbox}[
  enhanced, breakable, sharp corners,
  colback=gray!5, colframe=gray!50, boxrule=0.4pt,
  left=8pt, right=8pt, top=6pt, bottom=6pt
]
\small
\begin{verbatim}
# Site-level summaries: control-arm mean Y0bar, treatment-arm mean Y1bar,
# and covariate means X1bar, X2bar (averaged across both arms).
ctrl     <- subset(dat, Z == 0)
trt      <- subset(dat, Z == 1)
Y0bar    <- tapply(ctrl$Y, ctrl$j, mean)
Y1bar    <- tapply(trt$Y,  trt$j,  mean)
X1bar    <- tapply(dat$X1, dat$j,  mean)
X2bar    <- tapply(dat$X2, dat$j,  mean)
site_sum <- data.frame(j     = sort(unique(dat$j)),
                       Y0bar = as.numeric(Y0bar),
                       Y1bar = as.numeric(Y1bar),
                       X1bar = as.numeric(X1bar),
                       X2bar = as.numeric(X2bar))

# Attach pooled site-level covariate means back to the individual data
# (used by 2S-ME Step 2 centering).
dat$X1bar <- X1bar[dat$j]
dat$X2bar <- X2bar[dat$j]

# Control-arm covariate means per site (used by 2S-ME Step 1 centering).
X1bar_ctrl <- tapply(ctrl$X1, ctrl$j, mean)
X2bar_ctrl <- tapply(ctrl$X2, ctrl$j, mean)
\end{verbatim}
\end{tcolorbox}

\subsection*{Site-Level OLS with Summary Statistics (Site-Sum-OLS)}

A site-level baseline for settings where individual records are unavailable, and the least-squares counterpart of Site-Sum-Bayes. Each site's ITT effect is plugged in as the raw mean difference $\bar{Y}_{1j} - \bar{Y}_{0j}$ and regressed across sites on the control mean $\bar{Y}_{0j}$ and covariate means $\bar{\boldsymbol{X}}_j$. The residual is the LRE estimate at site $j$.

\begin{tcolorbox}[
  enhanced, breakable, sharp corners,
  colback=gray!5, colframe=gray!50, boxrule=0.4pt,
  left=8pt, right=8pt, top=6pt, bottom=6pt
]
\small
\begin{verbatim}
# Step 1 (Site-Sum-OLS): site-specific ITT estimate as the raw mean
# difference between arms, with no within-site covariate adjustment.
site_sum$itt_site_sum_ols <- site_sum$Y1bar - site_sum$Y0bar

# Step 2 (Site-Sum-OLS): regress ITT estimates on site-level moderators.
# Residuals are the LRE estimates.
step2_site_sum_ols <- lm(itt_site_sum_ols ~ Y0bar + X1bar + X2bar,
                         data = site_sum)
lre_site_sum_ols   <- as.numeric(resid(step2_site_sum_ols))
\end{verbatim}
\end{tcolorbox}

\subsection*{Plug-in OLS}

A two-step procedure that fits separate within-site OLS regressions in Step 1 to estimate site-specific ITT effects with within-site covariate adjustment, then regresses these on site-level summaries across sites in Step 2. The LRE estimate is the residual from Step 2. Each Step 1 regression enters treatment and covariates additively, with no treatment-by-covariate interaction. Under within-site randomization, omitting these interactions does not bias the site-specific ITT estimate, and affects only its precision. Step 1 loops over sites because each within-site regression uses only that site's rows; Site-Sum-OLS and 2S-ME avoid this loop because their Step 1 is a single global operation.

\begin{tcolorbox}[
  enhanced, breakable, sharp corners,
  colback=gray!5, colframe=gray!50, boxrule=0.4pt,
  left=8pt, right=8pt, top=6pt, bottom=6pt
]
\small
\begin{verbatim}
# Step 1 (Plug-in OLS): site-by-site OLS of outcome Y on treatment indicator Z and
# individual covariates X1, X2, with no Z-by-covariate interaction. The
# coefficient on Z is the site-specific ITT estimate.
itt_2sols <- by(dat, dat$j, function(d) {
  coef(lm(Y ~ Z + X1 + X2, data = d))[["Z"]]
})
site_sum$itt_2sols <- as.numeric(itt_2sols)

# Step 2 (Plug-in OLS): regress site-specific ITT estimates on site-level
# moderators. Residuals are the LRE estimates.
step2_2sols <- lm(itt_2sols ~ Y0bar + X1bar + X2bar, data = site_sum)
lre_2sols   <- as.numeric(resid(step2_2sols))
\end{verbatim}
\end{tcolorbox}

\subsection*{Two-Step Mixed-Effects (2S-ME)}
A two-step procedure adapted from \citet{Hong2026AJE}. Step 1 fits a control-arm random-intercept model. Site-level covariate means enter as predictors, and individual covariates centered at each site's control-arm mean enter for precision without shifting the site intercept. The random intercept $\tilde{\mu}_{0j}$ is the empirical Bayes estimate of each site's control-group mean after adjusting for its covariate composition. Restricting Step 1 to the control arm isolates local conditions in the absence of the program, untouched by treatment-effect heterogeneity. Step 2 fits a random-slope model on both arms, in which $\tilde{\mu}_{0j}$ and $\bar{\boldsymbol{X}}_j$ moderate the treatment indicator, and individual covariates centered at each site's pooled mean again enter for precision. The LRE estimate at site $j$ is the predicted site-specific treatment-indicator slope from this model. The individual covariate adjustment in both steps parallels the individual-data BHMs, keeping the comparison fair.

\begin{tcolorbox}[
  enhanced, breakable, sharp corners,
  colback=gray!5, colframe=gray!50, boxrule=0.4pt,
  left=8pt, right=8pt, top=6pt, bottom=6pt
]
\small
\begin{verbatim}
# Step 1 (2S-ME): control-only random-intercept model. Site covariate means
# adjust the control mean for composition; individual covariates centered at
# the control-arm site mean add within-site precision. The random intercept
# mu0_tilde is the EB estimate of each site's control-group mean.
ctrl$X1cc  <- ctrl$X1 - X1bar_ctrl[ctrl$j]   # centered at control-arm site mean
ctrl$X2cc  <- ctrl$X2 - X2bar_ctrl[ctrl$j]
step1_2sme <- lmer(Y ~ X1bar + X2bar + X1cc + X2cc + (1 | j), data = ctrl)
mu0_tilde  <- ranef(step1_2sme)$j[, "(Intercept)"]
dat$mu0_tilde <- mu0_tilde[dat$j]

# Step 2 (2S-ME): random-slope model on both arms. Z * mu0_tilde, Z * X1bar,
# Z * X2bar absorb the between-site treatment-effect variation explained by
# local conditions. Individual covariates centered at the pooled site mean
# add precision. The site-specific random Z slope is the LRE estimate.
dat$X1sc <- dat$X1 - X1bar[dat$j]            # centered at pooled site mean
dat$X2sc <- dat$X2 - X2bar[dat$j]
step2_2sme <- lmer(Y ~ Z * mu0_tilde + Z * X1bar + Z * X2bar +
                     X1sc + X2sc + (0 + Z | j),
                   data = dat)
lre_2sme <- as.numeric(ranef(step2_2sme)$j[, "Z"])
\end{verbatim}
\end{tcolorbox}

\subsection*{Constrained Bayes Rescaling (2S-ME-CB)}
A variant that applies the constrained Bayes rescaling of Louis (1984) and Ghosh (1992) to the 2S-ME empirical Bayes slopes. Empirical Bayes estimates are shrunk toward the grand mean, which improves per-site accuracy and gives the slopes a sample variance $\hat{\sigma}_{\mathrm{EB}}^2$ smaller than the model estimate of the between-site variance $\hat{\sigma}_\theta^2$. When the cross-site spread is itself of interest, the rescaling multiplies each slope's deviation from the mean by the inflation factor $\lambda = \hat{\sigma}_\theta / \hat{\sigma}_{\mathrm{EB}}$, so the rescaled estimates reproduce the between-site spread. Here $\hat{\sigma}_\theta^2$ estimates the between-site variance $\sigma_\theta^2$ in \eqref{eq:sigma_theta}, and $\hat{\sigma}_{\mathrm{EB}}^2$ is the sample variance of the empirical Bayes slopes. The rescaling reuses the Step 2 fit and changes only the spread of the point estimates, leaving their order unchanged.
\begin{tcolorbox}[
  enhanced, breakable, sharp corners,
  colback=gray!5, colframe=gray!50, boxrule=0.4pt,
  left=8pt, right=8pt, top=6pt, bottom=6pt
]
\small
\begin{verbatim}
# 2S-ME-CB: constrained Bayes rescaling of the 2S-ME EB slopes (lre_2sme).
# between-site variance
lre_var_between <- as.numeric(VarCorr(step2_2sme)$j["Z", "Z"])
# variance of the EB slopes
lre_var_eb      <- var(lre_2sme)
cb_scale        <- if (lre_var_eb > 0) sqrt(lre_var_between / lre_var_eb) else 1
# Inflate deviations so the spread matches the between-site variance.
lre_cb <- mean(lre_2sme) + cb_scale * (lre_2sme - mean(lre_2sme))
\end{verbatim}
\end{tcolorbox}
\subsection*{Triple-Goal Estimates (2S-ME-TG)}
A variant that constructs the set of site estimates to serve three goals together: the cross-site distribution, the rankings, and the per-site values (Shen \& Louis, 1998). Because the 2S-ME fit gives no joint posterior, we approximate each site's posterior by a normal distribution centered at its empirical Bayes slope, with standard deviation equal to the conditional standard deviation from the Step 2 fit. We then draw $M$ values per site. The support points are evenly spaced quantiles of the pooled draws, which estimate the cross-site distribution. Each site is then placed on the support point at its posterior expected rank. We set $M = 5000$. The support uses the $MJ$ pooled draws, and this $M$ keeps the Monte Carlo standard error of each pairwise rank comparison near $0.5/M^{1/2} \approx 0.007$.
\begin{tcolorbox}[
  enhanced, breakable, sharp corners,
  colback=gray!5, colframe=gray!50, boxrule=0.4pt,
  left=8pt, right=8pt, top=6pt, bottom=6pt
]
\small
\begin{verbatim}
# 2S-ME-TG: triple-goal estimates on the 2S-ME fit.
# Reuse the EB slopes (lre_2sme) and their conditional SDs from the Step 2 fit.
re_2sme <- ranef(step2_2sme, condVar = TRUE)$j
cond_sd <- sqrt(as.numeric(attr(re_2sme, "postVar")[1, 1, ]))
M       <- 5000
# Approximate each site's posterior by N(EB slope, conditional variance) and
# draw M values per site; columns are sites, rows are draws.
lre_draws <- matrix(rnorm(M * J, rep(lre_2sme, each = M), rep(cond_sd, each = M)),
                    nrow = M, ncol = J)
# Histogram target: evenly spaced quantiles of the pooled draws.
probs   <- (2 * seq_len(J) - 1) / (2 * J)
support <- quantile(as.vector(lre_draws), probs = probs, names = FALSE)
# Ranking target: posterior expected rank of each site.
exp_rank <- colMeans(t(apply(lre_draws, 1, rank, ties.method = "average")))
# Place each site on the support point at its expected rank.
lre_tg <- support[rank(exp_rank, ties.method = "first")]
\end{verbatim}
\end{tcolorbox}

\section{Descriptive Statistics of the NJCS Centers}
\label{app:njcs_desc}

Table~\ref{tab:njcs_desc} summarizes the 100 Job Corps centers in the analytic sample. Each row describes a center-level quantity computed from the individual records within a center.

\begin{table}[H]
\caption{Center-level descriptive statistics in the NJCS analytic sample ($J = 100$)}
\label{tab:njcs_desc}
\centering
\begin{threeparttable}
\small
\begin{tabular}{lrrrrr}
 & Mean & SD & Min & Median & Max \\
\midrule
Participants per center & 100.0 & 74.9 & 27 & 85.5 & 489 \\
Control participants per center & 39.3 & 29.4 & 10 & 33 & 179 \\
\addlinespace
Mean weekly earnings, control group (\$) & 196.4 & 40.5 & 100.9 & 195.9 & 285.3 \\
Mean weekly earnings, treatment group (\$) & 209.7 & 42.5 & 106.7 & 209.3 & 347.0 \\
\addlinespace
Mean age (years) & 18.73 & 0.48 & 17.56 & 18.72 & 19.79 \\
Proportion female & 0.39 & 0.18 & 0.00 & 0.40 & 0.94 \\
Mean log prior-year earnings & 4.45 & 0.71 & 2.22 & 4.48 & 6.39 \\
Mean months employed in the prior year & 3.49 & 0.73 & 1.21 & 3.50 & 5.38 \\
Proportion Black & 0.43 & 0.29 & 0.00 & 0.44 & 0.94 \\
Proportion Hispanic & 0.16 & 0.19 & 0.00 & 0.09 & 1.00 \\
\addlinespace
Proportion from superdense ZIP codes & 0.28 & 0.32 & 0.00 & 0.13 & 1.00 \\
Proportion from ZIP codes that are not dense & 0.43 & 0.28 & 0.00 & 0.44 & 1.00 \\
\end{tabular}
\begin{tablenotes}
\small
\item \textit{Note.} Each row summarizes a center-level quantity across the 100 centers. Means and proportions are computed within center from the individual records. Prior-year earnings enter as $\log(1 + \text{earnings})$.
\end{tablenotes}
\end{threeparttable}
\end{table}

\section{Additional Simulation Results}
\label{app:extra_results}

\begin{figure}[!htbp]
    \centering
    \includegraphics[width=\linewidth]{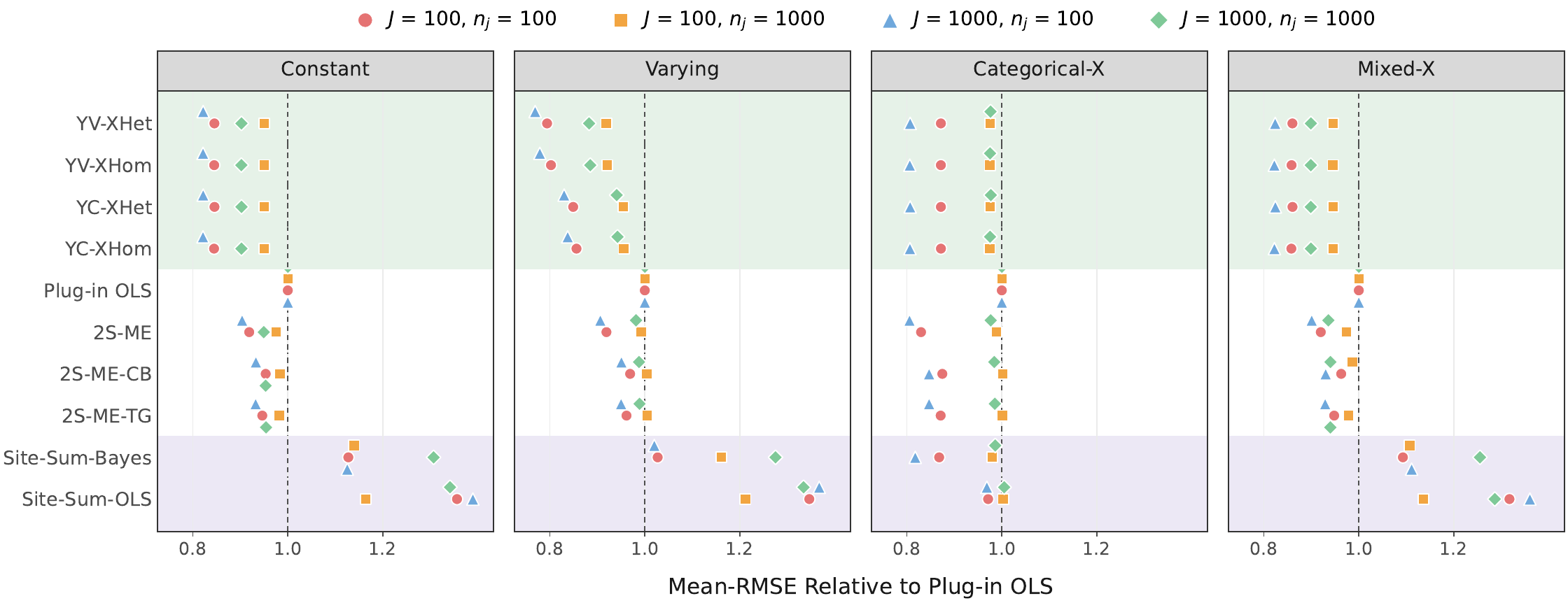}
    \caption{Mean-RMSE of $\hat{\theta}_j$ relative to Plug-in OLS for all ten methods, by DGP and design cell. The dashed line marks a ratio of one. Colors, shapes, and bands are as in Figure~\ref{fig:rmse_rel}.}
    \label{fig:rmse_all}
\end{figure}

\begin{figure}[!htbp]
    \centering
    \includegraphics[width=\linewidth]{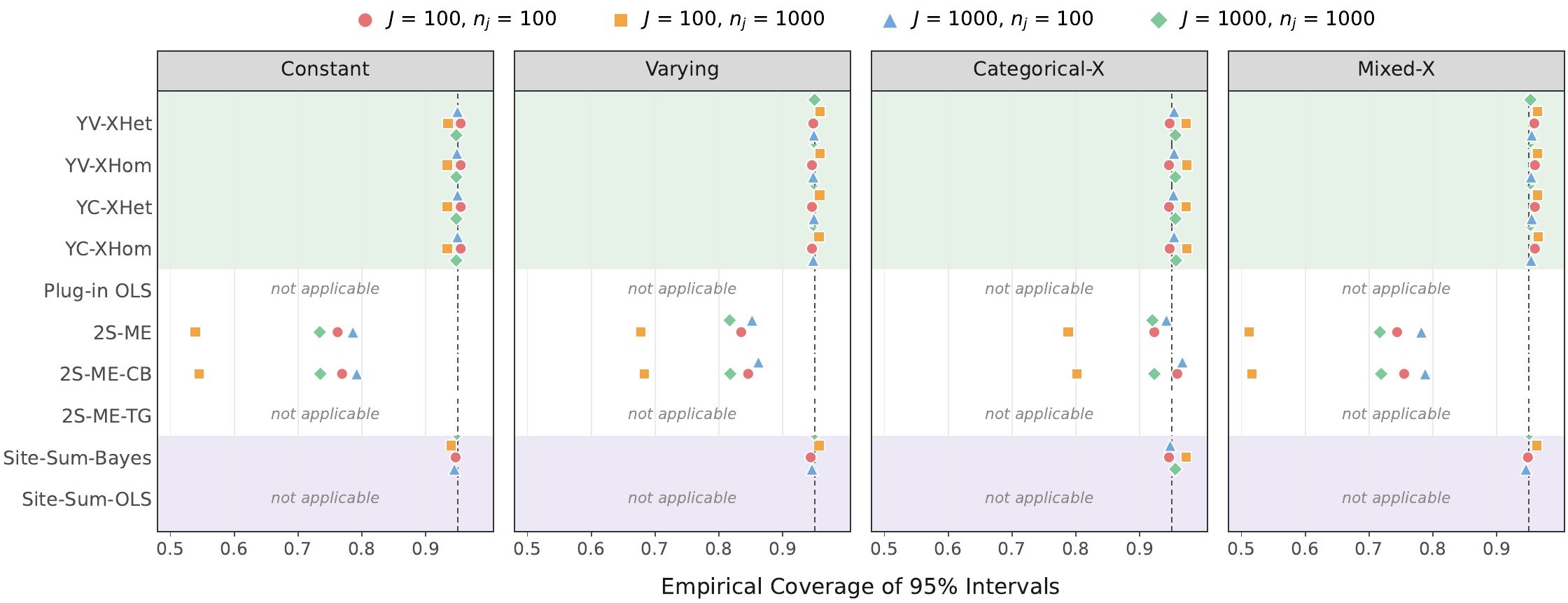}
    \caption{Empirical coverage of 95\% intervals for $\theta_j$ for all ten methods, by DGP and design cell. The dashed line marks the nominal 0.95. Coverage is not applicable to the three methods that return point estimates only.}
    \label{fig:coverage_all}
\end{figure}

\begin{figure}[!htbp]
    \centering
    \includegraphics[width=\linewidth]{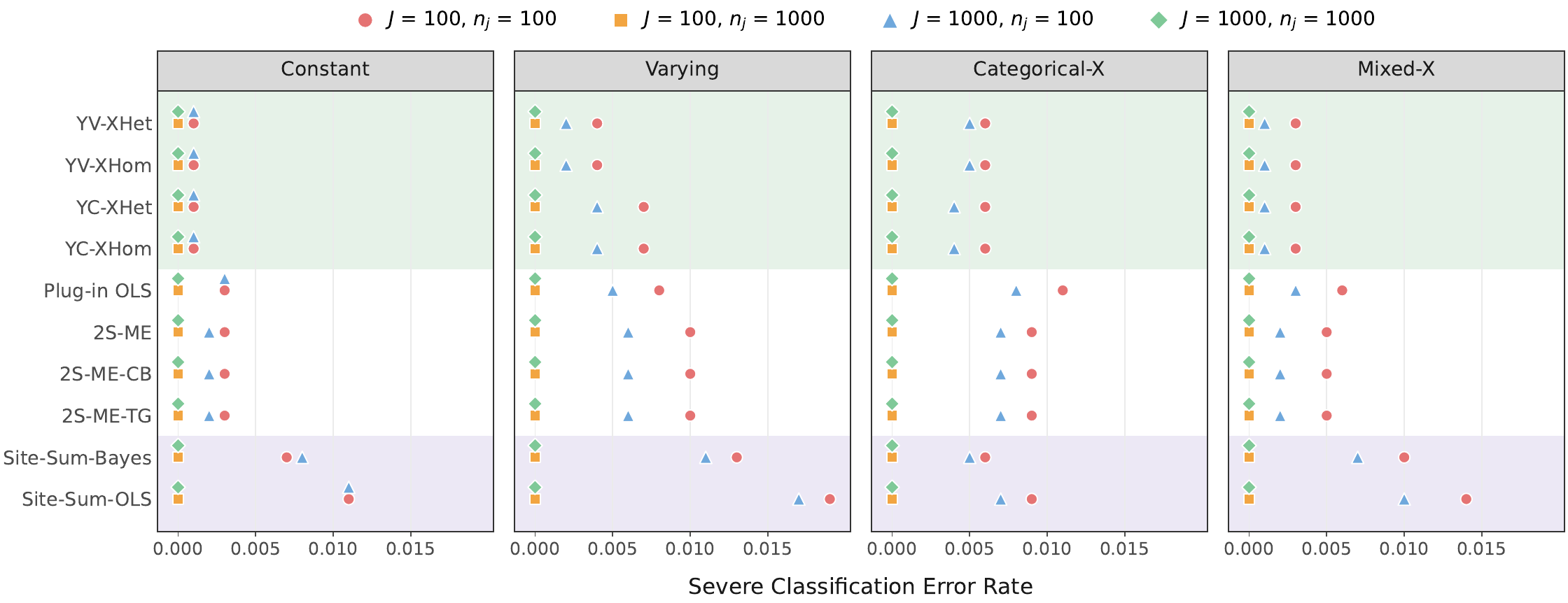}
    \caption{Severe classification error (SCE) rate for all ten methods, by DGP and design cell.}
    \label{fig:sce_all}
\end{figure}

\begin{figure}[!htbp]
    \centering
    \includegraphics[width=\linewidth]{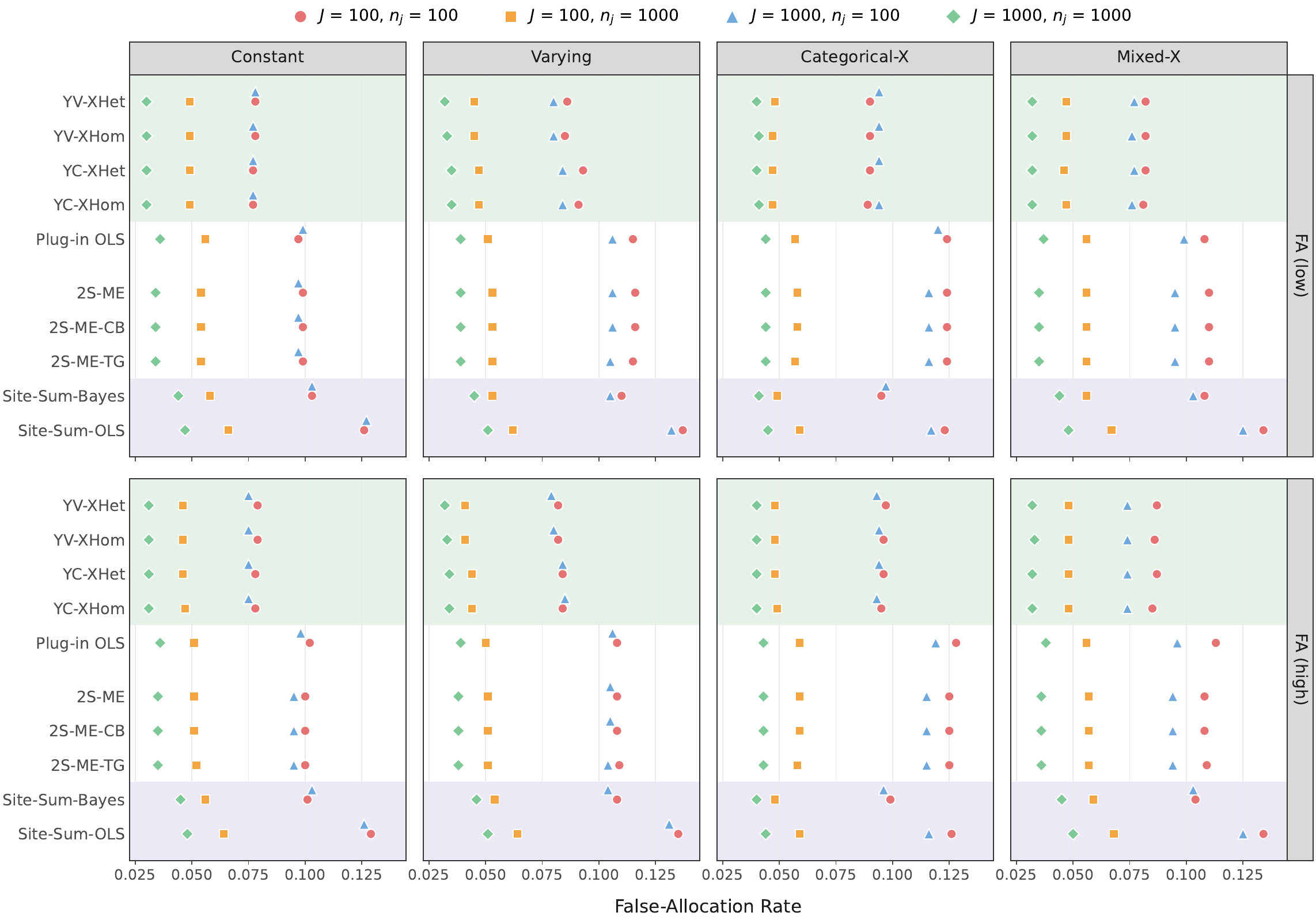}
    \caption{False-allocation rates for all ten methods, by DGP and design cell. The upper row shows $\mathrm{FA}^{\mathrm{low}}$ and the lower row $\mathrm{FA}^{\mathrm{high}}$.}
    \label{fig:fa_all}
\end{figure}

Figures~\ref{fig:rmse_all} to~\ref{fig:fa_all} repeat the four figures of Section~\ref{sec:results} for all ten methods. They add the two individual-data BHMs of Appendix~\ref{app:other_models} and the two adjustments of the 2S-ME estimates described in Appendix~\ref{app:comparison_methods}.

\textit{Additional individual-data BHMs.} YV-XHom and YC-XHet track the two main-text models closely on every measure. Under the Varying DGP at $n_j = 100$, the two models with site-varying slopes (YV-XHet, YV-XHom) lie to the left of the two with constant slopes (YC-XHet, YC-XHom) in Figure~\ref{fig:rmse_all}. Holding the slopes and residual variances constant therefore costs a little accuracy, whereas sharing the covariate covariance matrix costs almost nothing on any measure.

\begin{figure}[!htbp]
    \centering
    \includegraphics[width=0.9\linewidth]{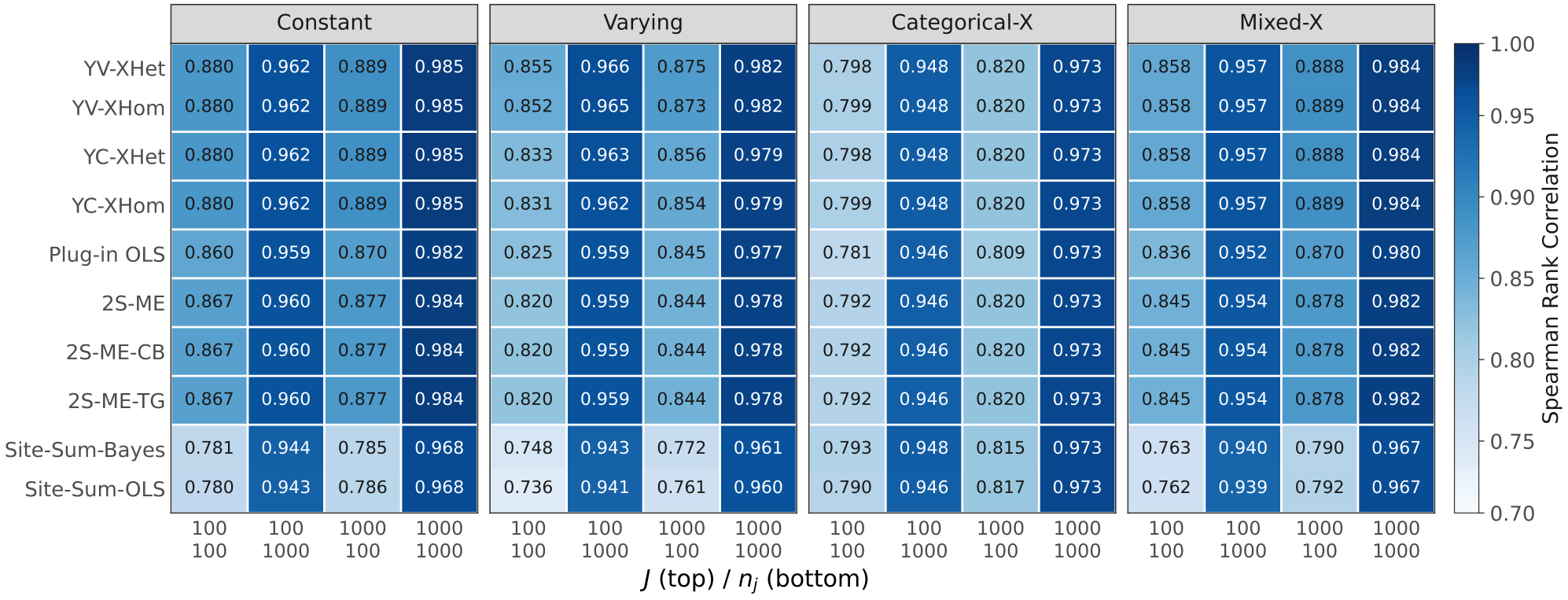}
    \caption{Spearman rank correlation between $\hat{\theta}_j$ and $\theta_j$ by method, DGP, and design cell, for the Constant, Varying, Categorical-$\boldsymbol{X}$, and Mixed-$\boldsymbol{X}$ DGPs. Columns are labeled by $J$ and $n_j$.}
    \label{fig:spearman_heat}
\end{figure}

\textit{Adjustments of 2S-ME.} We expected the constrained Bayes adjustment to preserve the order of the 2S-ME estimates, so that its rankings match 2S-ME while its rescaling enlarges per-site error. We expected the triple-goal adjustment to recover the cross-site spread and the rankings more closely, at some cost to per-site accuracy. In Figure~\ref{fig:rmse_all}, 2S-ME-CB and 2S-ME-TG appear slightly to the right of 2S-ME. Both take the 2S-ME estimates as given and adjust them, by rescaling or by re-positioning at expected ranks, without re-estimating anything, so they cannot reduce the error in those estimates and in general add to it. 2S-ME-CB keeps the 2S-ME rankings and tier rates, as expected. Its interval is centered at the rescaled estimate, with the conditional standard deviation multiplied by the same factor, so its coverage is close to that of 2S-ME (Figure~\ref{fig:coverage_all}). 2S-ME-TG places each site at an expected rank and produces no interval. Contrary to our expectation, its tier rates (Figures~\ref{fig:sce_all} and~\ref{fig:fa_all}) and rank correlations (Figure~\ref{fig:spearman_heat}) match those of 2S-ME. The triple-goal adjustment improves rankings when the empirical Bayes estimates are shrunk unevenly across sites, so that shrinkage reorders them. With 100 or more sites of equal size, every site is shrunk by about the same factor, the 2S-ME ranking is already the triple-goal ranking, and the adjustment has nothing left to do.

\begin{figure}[!htbp]
    \centering
    \includegraphics[width=0.9\linewidth]{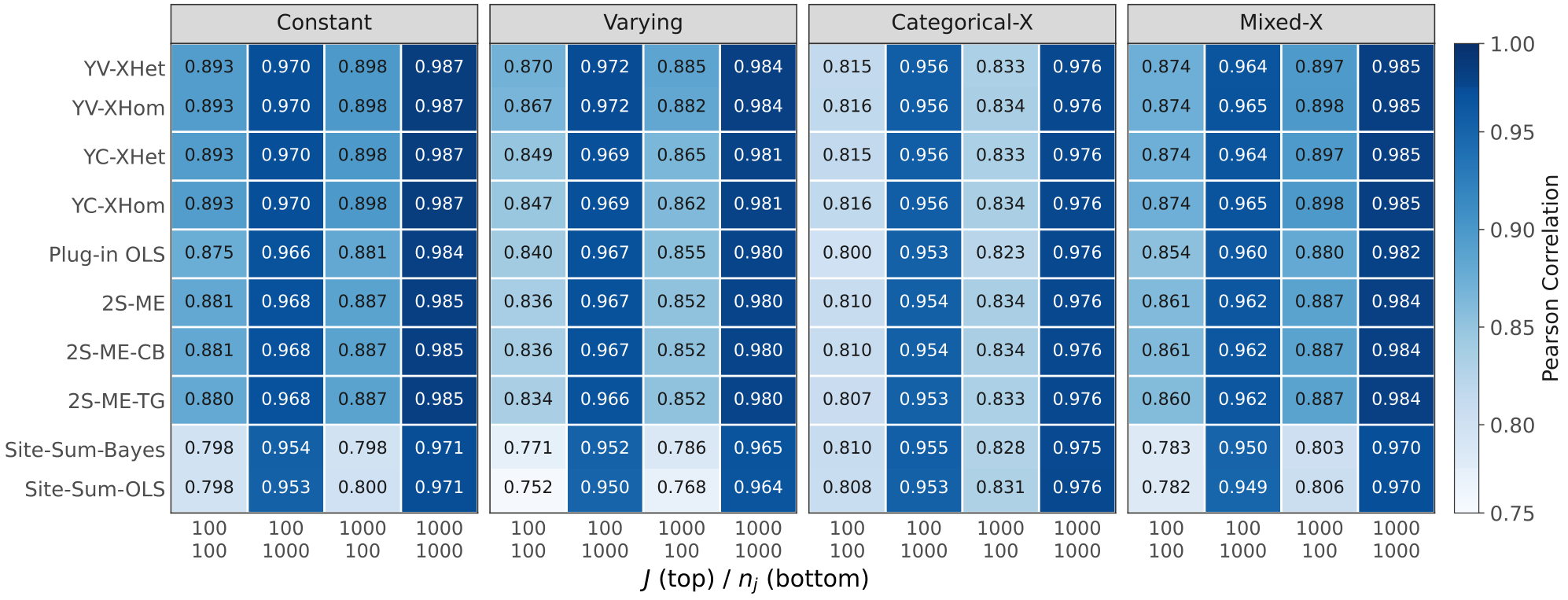}
    \caption{Pearson correlation between $\hat{\theta}_j$ and $\theta_j$ by method, DGP, and design cell, for the Constant, Varying, Categorical-$\boldsymbol{X}$, and Mixed-$\boldsymbol{X}$ DGPs. Columns are labeled by $J$ and $n_j$.}
    \label{fig:pearson_heat}
\end{figure}

Figures~\ref{fig:pearson_heat} and~\ref{fig:spearman_heat} show Pearson and Spearman correlations between estimated and true site-specific LREs across the four DGPs in the main text, with methods ordered as in Figure~\ref{fig:rmse_all}. Results for the Varying-Slope DGP appear in Appendix~\ref{app:varying}. Both correlations increase with within-site sample size but vary little with the number of sites.

\section{Results under the Varying-Slope DGP}
\label{app:varying}

The Varying-Slope DGP draws the within-site slopes and residual variances from hierarchical distributions and holds the covariate covariance at $I_2$, so YV-XHom is the matching model (Appendix~\ref{app:dgp_models}). Tables~\ref{tab:var_rmse} to~\ref{tab:var_fahigh} report the four measures of Section~\ref{sec:results} in the same method order. Tables~\ref{tab:var_pearson} and~\ref{tab:var_spearman} report the Pearson and Spearman correlations between the estimated and true site-specific LREs. Every pattern in Section~\ref{sec:results} reappears, and on each measure the Varying-Slope DGP sits between the Constant and Varying DGPs.

\begin{table}[H]
\caption{Mean-RMSE of $\hat{\theta}_j$ relative to Plug-in OLS: Varying-Slope DGP}
\label{tab:var_rmse}
\centering
\begin{threeparttable}
\small
\begin{tabular}{lcccc}
 & $J{=}100$ & $J{=}100$ & $J{=}1000$ & $J{=}1000$ \\
Method & $n_j{=}100$ & $n_j{=}1000$ & $n_j{=}100$ & $n_j{=}1000$ \\
\midrule
YV-XHet & 0.816 & 0.939 & 0.789 & 0.890 \\
YV-XHom & 0.816 & 0.939 & 0.789 & 0.890 \\
YC-XHet & 0.866 & 0.969 & 0.843 & 0.945 \\
YC-XHom & 0.865 & 0.969 & 0.843 & 0.945 \\
\addlinespace
Plug-in OLS & 1.000 & 1.000 & 1.000 & 1.000 \\
2S-ME & 0.923 & 0.991 & 0.906 & 0.984 \\
2S-ME-CB & 0.971 & 1.001 & 0.948 & 0.990 \\
2S-ME-TG & 0.963 & 1.002 & 0.948 & 0.991 \\
\addlinespace
Site-Sum-Bayes & 1.059 & 1.161 & 1.044 & 1.281 \\
Site-Sum-OLS & 1.339 & 1.189 & 1.349 & 1.329 \\
\end{tabular}
\begin{tablenotes}
\small
\item \textit{Note.} Ratio of each method's Mean-RMSE to that of Plug-in OLS in the same cell, averaged over 100 replications; values below one indicate greater accuracy than Plug-in OLS.
\end{tablenotes}
\end{threeparttable}
\end{table}

\begin{table}[H]
\caption{Empirical coverage of 95\% intervals for $\theta_j$: Varying-Slope DGP}
\label{tab:var_cov}
\centering
\begin{threeparttable}
\small
\begin{tabular}{lcccc}
 & $J{=}100$ & $J{=}100$ & $J{=}1000$ & $J{=}1000$ \\
Method & $n_j{=}100$ & $n_j{=}1000$ & $n_j{=}100$ & $n_j{=}1000$ \\
\midrule
YV-XHet & 0.954 & 0.953 & 0.950 & 0.951 \\
YV-XHom & 0.955 & 0.953 & 0.949 & 0.950 \\
YC-XHet & 0.952 & 0.950 & 0.948 & 0.949 \\
YC-XHom & 0.952 & 0.950 & 0.948 & 0.949 \\
\addlinespace
Plug-in OLS & -- & -- & -- & -- \\
2S-ME & 0.833 & 0.678 & 0.851 & 0.815 \\
2S-ME-CB & 0.849 & 0.684 & 0.860 & 0.817 \\
2S-ME-TG & -- & -- & -- & -- \\
\addlinespace
Site-Sum-Bayes & 0.946 & 0.953 & 0.946 & 0.950 \\
Site-Sum-OLS & -- & -- & -- & -- \\
\end{tabular}
\begin{tablenotes}
\small
\item \textit{Note.} Nominal value is 0.95. Credible intervals for the Bayesian models; normal-approximation intervals for 2S-ME and 2S-ME-CB. Coverage is not applicable to methods marked --, which return point estimates only.
\end{tablenotes}
\end{threeparttable}
\end{table}

\begin{table}[H]
\caption{Severe classification error (SCE) rate: Varying-Slope DGP}
\label{tab:var_sce}
\centering
\begin{threeparttable}
\small
\begin{tabular}{lcccc}
 & $J{=}100$ & $J{=}100$ & $J{=}1000$ & $J{=}1000$ \\
Method & $n_j{=}100$ & $n_j{=}1000$ & $n_j{=}100$ & $n_j{=}1000$ \\
\midrule
YV-XHet & 0.003 & 0.000 & 0.002 & 0.000 \\
YV-XHom & 0.003 & 0.000 & 0.002 & 0.000 \\
YC-XHet & 0.003 & 0.000 & 0.003 & 0.000 \\
YC-XHom & 0.003 & 0.000 & 0.003 & 0.000 \\
\addlinespace
Plug-in OLS & 0.005 & 0.000 & 0.004 & 0.000 \\
2S-ME & 0.006 & 0.000 & 0.005 & 0.000 \\
2S-ME-CB & 0.006 & 0.000 & 0.005 & 0.000 \\
2S-ME-TG & 0.006 & 0.000 & 0.005 & 0.000 \\
\addlinespace
Site-Sum-Bayes & 0.009 & 0.000 & 0.009 & 0.000 \\
Site-Sum-OLS & 0.014 & 0.000 & 0.013 & 0.000 \\
\end{tabular}
\begin{tablenotes}
\small
\item \textit{Note.} Share of sites assigned to the tail opposite their true tail, averaged over replications.
\end{tablenotes}
\end{threeparttable}
\end{table}

\begin{table}[H]
\caption{False-allocation rate into the low tier, $\mathrm{FA}^{\mathrm{low}}$: Varying-Slope DGP}
\label{tab:var_falow}
\centering
\begin{threeparttable}
\small
\begin{tabular}{lcccc}
 & $J{=}100$ & $J{=}100$ & $J{=}1000$ & $J{=}1000$ \\
Method & $n_j{=}100$ & $n_j{=}1000$ & $n_j{=}100$ & $n_j{=}1000$ \\
\midrule
YV-XHet & 0.080 & 0.043 & 0.079 & 0.032 \\
YV-XHom & 0.082 & 0.043 & 0.080 & 0.032 \\
YC-XHet & 0.086 & 0.045 & 0.083 & 0.034 \\
YC-XHom & 0.086 & 0.045 & 0.083 & 0.034 \\
\addlinespace
Plug-in OLS & 0.105 & 0.050 & 0.103 & 0.038 \\
2S-ME & 0.107 & 0.050 & 0.103 & 0.037 \\
2S-ME-CB & 0.107 & 0.050 & 0.103 & 0.037 \\
2S-ME-TG & 0.107 & 0.050 & 0.103 & 0.037 \\
\addlinespace
Site-Sum-Bayes & 0.107 & 0.052 & 0.104 & 0.044 \\
Site-Sum-OLS & 0.136 & 0.060 & 0.129 & 0.049 \\
\end{tabular}
\begin{tablenotes}
\small
\item \textit{Note.} Share of truly non-low sites assigned to the low tier.
\end{tablenotes}
\end{threeparttable}
\end{table}

\begin{table}[H]
\caption{False-allocation rate into the high tier, $\mathrm{FA}^{\mathrm{high}}$: Varying-Slope DGP}
\label{tab:var_fahigh}
\centering
\begin{threeparttable}
\small
\begin{tabular}{lcccc}
 & $J{=}100$ & $J{=}100$ & $J{=}1000$ & $J{=}1000$ \\
Method & $n_j{=}100$ & $n_j{=}1000$ & $n_j{=}100$ & $n_j{=}1000$ \\
\midrule
YV-XHet & 0.081 & 0.043 & 0.079 & 0.032 \\
YV-XHom & 0.080 & 0.043 & 0.078 & 0.032 \\
YC-XHet & 0.085 & 0.044 & 0.082 & 0.034 \\
YC-XHom & 0.084 & 0.043 & 0.082 & 0.034 \\
\addlinespace
Plug-in OLS & 0.109 & 0.053 & 0.105 & 0.038 \\
2S-ME & 0.107 & 0.049 & 0.102 & 0.038 \\
2S-ME-CB & 0.107 & 0.049 & 0.102 & 0.038 \\
2S-ME-TG & 0.107 & 0.050 & 0.102 & 0.038 \\
\addlinespace
Site-Sum-Bayes & 0.106 & 0.056 & 0.105 & 0.045 \\
Site-Sum-OLS & 0.130 & 0.062 & 0.128 & 0.050 \\
\end{tabular}
\begin{tablenotes}
\small
\item \textit{Note.} Share of truly non-high sites assigned to the high tier.
\end{tablenotes}
\end{threeparttable}
\end{table}

\begin{table}[H]
\caption{Pearson correlation between $\hat{\theta}_j$ and $\theta_j$: Varying-Slope DGP}
\label{tab:var_pearson}
\centering
\begin{threeparttable}
\small
\begin{tabular}{lcccc}
 & $J{=}100$ & $J{=}100$ & $J{=}1000$ & $J{=}1000$ \\
Method & $n_j{=}100$ & $n_j{=}1000$ & $n_j{=}100$ & $n_j{=}1000$ \\
\midrule
YV-XHet & 0.878 & 0.972 & 0.888 & 0.985 \\
YV-XHom & 0.878 & 0.972 & 0.888 & 0.985 \\
YC-XHet & 0.861 & 0.970 & 0.871 & 0.983 \\
YC-XHom & 0.861 & 0.970 & 0.871 & 0.983 \\
\addlinespace
Plug-in OLS & 0.855 & 0.968 & 0.862 & 0.981 \\
2S-ME & 0.852 & 0.969 & 0.862 & 0.981 \\
2S-ME-CB & 0.852 & 0.969 & 0.862 & 0.981 \\
2S-ME-TG & 0.851 & 0.968 & 0.861 & 0.981 \\
\addlinespace
Site-Sum-Bayes & 0.782 & 0.955 & 0.793 & 0.968 \\
Site-Sum-OLS & 0.774 & 0.955 & 0.785 & 0.967 \\
\end{tabular}
\begin{tablenotes}
\small
\item \textit{Note.} Averaged over 100 replications.
\end{tablenotes}
\end{threeparttable}
\end{table}

\begin{table}[H]
\caption{Spearman rank correlation between $\hat{\theta}_j$ and $\theta_j$: Varying-Slope DGP}
\label{tab:var_spearman}
\centering
\begin{threeparttable}
\small
\begin{tabular}{lcccc}
 & $J{=}100$ & $J{=}100$ & $J{=}1000$ & $J{=}1000$ \\
Method & $n_j{=}100$ & $n_j{=}1000$ & $n_j{=}100$ & $n_j{=}1000$ \\
\midrule
YV-XHet & 0.865 & 0.966 & 0.878 & 0.983 \\
YV-XHom & 0.865 & 0.966 & 0.878 & 0.983 \\
YC-XHet & 0.847 & 0.964 & 0.861 & 0.981 \\
YC-XHom & 0.847 & 0.964 & 0.861 & 0.981 \\
\addlinespace
Plug-in OLS & 0.839 & 0.962 & 0.852 & 0.979 \\
2S-ME & 0.839 & 0.962 & 0.852 & 0.979 \\
2S-ME-CB & 0.839 & 0.962 & 0.852 & 0.979 \\
2S-ME-TG & 0.839 & 0.962 & 0.852 & 0.979 \\
\addlinespace
Site-Sum-Bayes & 0.766 & 0.947 & 0.779 & 0.964 \\
Site-Sum-OLS & 0.760 & 0.947 & 0.774 & 0.964 \\
\end{tabular}
\begin{tablenotes}
\small
\item \textit{Note.} Averaged over 100 replications.
\end{tablenotes}
\end{threeparttable}
\end{table}

\clearpage
\begin{center}\textbf{Supplementary References}\end{center}

\begin{list}{}{\setlength{\leftmargin}{0.5in}\setlength{\itemindent}{-0.5in}\setlength{\itemsep}{0pt}\setlength{\parsep}{0pt}}
\item Betancourt, M. (2017). \textit{A conceptual introduction to Hamiltonian Monte Carlo}. arXiv. \url{https://arxiv.org/abs/1701.02434}
\item Betancourt, M., \& Girolami, M. (2015). Hamiltonian Monte Carlo for hierarchical models. In S. K. Upadhyay, U. Singh, D. K. Dey, \& A. Loganathan (Eds.), \textit{Current trends in Bayesian methodology with applications} (pp. 79--101). Chapman and Hall/CRC.
\item Ghosh, M. (1992). Constrained Bayes estimation with applications. \textit{Journal of the American Statistical Association}, \textit{87}(418), 533--540. doi: 10.1080/01621459.1992.10475236
\item Louis, T. A. (1984). Estimating a population of parameter values using Bayes and empirical Bayes methods. \textit{Journal of the American Statistical Association}, \textit{79}(386), 393--398. doi: 10.1080/01621459.1984.10478062
\item Papaspiliopoulos, O., Roberts, G. O., \& Sk\"{o}ld, M. (2007). A general framework for the parametrization of hierarchical models. \textit{Statistical Science}, \textit{22}(1), 59--73. doi: 10.1214/088342307000000014
\item Polson, N. G., \& Scott, J. G. (2012). On the half-Cauchy prior for a global scale parameter. \textit{Bayesian Analysis}, \textit{7}(4), 887--902. doi: 10.1214/12-BA730
\item Shen, W., \& Louis, T. A. (1998). Triple-goal estimates in two-stage hierarchical models. \textit{Journal of the Royal Statistical Society: Series B (Statistical Methodology)}, \textit{60}(2), 455--471. doi: 10.1111/1467-9868.00135
\item Stan Development Team. (2024). \textit{Prior choice recommendations}. Stan Wiki. \url{https://github.com/stan-dev/stan/wiki/Prior-Choice-Recommendations}
\end{list}

\end{document}